\documentclass[]{aa}
\usepackage[T1]{fontenc}
\usepackage[utf8]{inputenc}

\usepackage{graphicx}
\usepackage{natbib}
\usepackage{ulem}
\usepackage{textcomp}
\usepackage{gensymb}
\usepackage{longtable}
\usepackage{threeparttable}
\usepackage{multicol}
\usepackage{multirow}
\usepackage{float}
\usepackage{todonotes}
\usepackage[Symbol]{upgreek}
\usepackage{array}
\usepackage{amsmath}
\usepackage{etoolbox}
\usepackage{txfonts}
\usepackage{url}
\usepackage{xcolor}

\usepackage[most]{tcolorbox}
\usepackage{xcolor}
\newcommand{\erosita}{{\small eROSITA}}
\newcommand{\apec}{{\small APEC} }
\newcommand{\healpix}{{\small HEALPix }}

\begin{document}

\title{The warm-hot circumgalactic medium of the Milky Way as seen by eROSITA}

\author{N. Locatelli\inst{1,2}, G. Ponti\inst{2,1}, X. Zheng\inst{1}, A. Merloni\inst{1}, W. Becker\inst{1}, J. Comparat\inst{1}, K. Dennerl\inst{1}, M. J. Freyberg\inst{1}, M. Sasaki\inst{3}, M. C. H. Yeung\inst{1} }

\offprints{%
 E-mail: nicola.locatelli@inaf.it }
\institute{
Max-Planck-Institut f\"ur Extraterrestrische Physik (MPE), Giessenbachstrasse 1, 85748 Garching bei M\"unchen, Germany
\and INAF - Osservatorio Astronomico di Brera, via E. Bianchi 46, 23807 Merate (LC), Italy
\and Dr. Karl Remeis Observatory, Erlangen Centre for Astroparticle Physics, Friedrich-Alexander-Universit\"at Erlangen-N\"urnberg, Sternwartstra{\ss}e 7, 96049 Bamberg, Germany.
}

\authorrunning{N. Locatelli et al.}
\titlerunning{eROSITA view of the MW hot CGM}

\date{Accepted ???. Received ???; in original form ???}

\abstract{
The first all-sky maps of the diffuse emission of high ionization lines observed in X-rays by SRG/\erosita\, provide an excellent probe for the study of the warm-hot phase ($T\sim 10^6$ K) of the circumgalactic medium (CGM) of the Milky Way (MW). 
In this work we analyse the O VIII line detected in the first eROSITA All-Sky Survey data (eRASS1). 
We fit a sky map made in a narrow energy bin around this line, with physical emission models embedded in a 3D geometry to constrain the density distribution of the warm-hot gas around our Galaxy, with a focus on mid and high (absolute) Galactic latitudes.
By masking out the \erosita\, bubbles and other bright extended foreground sources, we find that an oblate geometry of the warm-hot gas ($T\equiv 0.15-0.17$ keV), flattened around the Galactic disk with scale height $z_h\sim 1-3$ kpc, best describes the eRASS1 O VIII map, with most of the observed emission resulting to be produced within a few kpc from the Sun.
The additional presence of a large scale warm-hot spherical halo, while providing a minor contribute to the X-ray emission, accounts for the high O VII absorption column densities detected with XMM-{\it Newton}, as well as most of the baryon budget of the CGM of the MW. The \erosita{} data carry the largest amount of information and detail of O VIII CGM intensities to date, allowing for a significant reduction of the statistical uncertainties of the inferred physical parameters.
}
\maketitle

\label{firstpage}

\section{Introduction}

The largest contribution to the gas mass budget of galaxies is expected to be retained in a hot phase in the halo which extents over scales that are comparable to their virial radius $R_{\rm vir}$, with temperature $T\sim 10^5-10^{7}$ K \citep{1978MNRAS.183..341W, 2017ARA&A..55..389T}.
The presence of these hot gas halos results from the infall of intergalactic material onto the spines and nodes of the dark matter structure of the Universe (i.e. the cosmic web), from the $\sim$Mpc scales down onto the smaller scale peaks ($\sim$100 kpc) corresponding to the galactic dark matter halos. The infalling gas bulk motions and collisions power stationary shock waves at the boundaries of the potential wells. These stationary shocks compress and heat up the gas inside the well, up to a temperature similar (but not necessarily equal, see \citealt{2022arXiv220609925L}) to the virial temperature $T_{\rm vir} \propto R_{\rm vir} / M_{\rm vir}\sim 10^5-10^7$ K usually computed at $R_{\rm vir}\equiv R_{200}$ \citep[see][and references therein]{2018MNRAS.481..835O, 2018MNRAS.477..450N}. 
The medium enclosed within this radius and found outside the stellar disk of a galaxy is usually defined as the circumgalactic medium (CGM).

In this warm-hot gas phase of the CGM, collisions between the gas atoms dominate the energy exchange between them and in turn the overall ionisation of the atomic species. The collisional ionisation equilibrium hypothesis thus provides a theoretical framework to compute the expected brightness of the ionisation lines of the warm-hot gas phase \citep{2001ApJ...556L..91S}.
Such a phase has been observed as all-sky diffuse X-ray emission since about three decades already  \citep[e.g.][based on ROSAT all-sky X-ray survey data]{2000ApJ...543..195K}.
High ionisation lines of species like C IV, O VII, O VIII or Ne IX are the most common states and their presence has been confirmed around external galaxies by several independent probes:
via the absorption features they produce along the lines of sight towards bright active galactic nuclei \citep{2012ApJ...756L...8G, 2013ApJ...770..118M, 2023arXiv230204247N, 2019ApJ...887..257D, 2019ApJ...882L..23D}; 
via their associated emission lines studies (\citealt{1985ApJ...293..102F, 2001MNRAS.328..461O, 2009ApJ...690..143Y, 2010ApJS..187..388H, 2013ApJ...772...97B, 2015ApJ...800...14M, 2016ApJ...826..167G, 2017ApJ...835...52F}, although see potential biases pointed out by \citealt{2020ApJ...896..143Z});
by detecting diffuse emission around external galaxies \citep{2004ApJS..151..193S, 2006A&A...448...43T, 2017ApJS..233...20L, 2018ApJ...866..126H};
via stacking experiments over a large sample of distant galaxies, revealing the presence of a layer of hot gas distributed closely within and around galactic stellar discs \citep{2015MNRAS.449.3806A, 2022A&A...666A.156C, 2022ApJ...936L..15C}. 
In the context of the MW, the spectral evidence of both warm-hot $T\simeq 0.2-0.3$ keV and hot $T\simeq 0.7-1$ keV gas phases have been reported \citep{2009PASJ...61..805Y, 2012ApJ...756L...8G, 2018ApJ...862...34N, 2019ApJ...887..257D, 2019ApJ...882L..23D, 2020NatAs...4.1072K, 2022arXiv220807863B, 2022arXiv221003133P, 2022ApJ...936...72B} and similar components have also been recently associated to the Large Magellanic Cloud \citep{2021AJ....161...57G}.

The same set of evidences however, may also suggest the alternative scenario in which the heated gas is expelled from the stellar disc from the explosions of supernovae, by mechanical or radiative energy feedback. The gas outflowing from the stellar disk may then cool, precipitate and fall back onto the disc creating a re-cycle of the gas powering new episodes of star-formation  \citep{1976ApJ...205..762S, 1980ApJ...236..577B}. 
Given the sensitivity of X-ray experiments to particle density, usually increasing towards the inner portions of galactic halos, this scenario is a complementary alternative to the gravitational infall in providing an explanation for the presence of a hot gas phase around galaxies.

The current picture for the MW indicates the presence of both types of scenarios (gas accreted from the large scale environment vs. outflowing from the Galactic disc), with a component distributed in a disk-like geometry extending $\sim$ a few kpc above and below the Galactic plane producing most of the observed X-ray CGM emission, while a large scale ($\sim 100-300$ kpc) halo is present but provides a minor contribution to the emission \citep{2018ApJ...862...34N, 2020NatAs...4.1072K, 2019ApJ...880...89Q, 2022ApJ...936...72B}. The halo component however, would contain most of the mass associated to the hot gas phase due to its huge volume. 

A crucial aspect in studies of the diffuse emission of the MW is to cover large portions of the sky with sufficient spatial resolution to discriminate sources morphology and sufficient spectral resolution to distinguish emission components. The sky fraction sampled by instruments with good spatial and spectral resolution (e.g. XMM-{\it Newton, Chandra}) is small due to the relatively small field of view (compared to $4\pi$ sr). In this respect, the {\it ROSAT} mission \citep{1990Ap&SS.171..207S, 1997ApJ...485..125S} operational in the '90s set a milestone producing the first all-sky X-ray map. The spectral resolution of {\it ROSAT}, providing five broad bands from 0.1 to 2.4 keV, prevented however studies of single emission lines and an easy identification of the different sources of diffuse emission in a given energy band.
More recently the HaloSat instrument provided a larger coverage at high Galactic latitudes with $10\deg$ spatial resolution and relatively good spectral resolution (85 eV at 0.68 keV, \citealt{2019ApJ...884..162K}).
The best figure of merit exploiting a high survey speed and a sufficient spectral resolution, combined with a high spatial resolution, has finally been reached by the extended ROentgen Survey with an Imaging Telescope Array (\erosita) instrument onboard the Spectrum-Roentgen-Gamma (SRG) space observatory, launched in July 2019. 
\erosita\, is a space X-ray telescope featuring a large effective area from 0.2 to 8 keV (comparable to that of XMM-{\it Newton} in the 0.3-2 keV band), in combination with a large field of view ($\sim 1$ deg$^2$), high spatial resolution ($\sim 30$ arcsec) and instrumental energy resolution of $\sim 80$ eV at 1 keV \citep{2012arXiv1209.3114M, 2021A&A...656A.132S, 2021A&A...647A...1P}.
These features combined, allow for the first time to break down the soft X-ray background (0.2-1 keV) into its components, including the hot CGM of the MW, in both spectra and high-resolved images.

In this paper, we present the analysis of the first ever O VIII half-sky\footnote{the Western half of the sky at $359.9442 > l >179.9442$ degrees (great circle over galactic poles and Sgr A*). These are proprietary data of the \erosita{}\_DE consortium.} 
emission lines maps (Zheng et al., in preparation), aimed at constraining the density distribution and overall geometry of the warm-hot CGM of the MW.
The spectral analysis performed on the \erosita\ Final Equatorial Depth Survey (eFEDS, \citealt{2022A&A...661A...1B}), a deep $\sim 140\, \rm deg^2$ field at moderate Galactic latitudes ($20^\circ<b<40^\circ$, \citealt[][and references therein]{2022arXiv221003133P}), provides us with information on the temperature and metal content of the detected CGM phases.
By adopting these information, we can derive the X-ray emission of simple 3D geometrical models for the CGM density, and look for the one best describing the narrow-band map of the detected O VIII emission line.

In Sec.~\ref{sec:data} we briefly summarize the data reduction extensively presented in Zheng et al., submitted; in Sec.~\ref{sec:model} we describe the components used to describe the line emission and the method used to fit the CGM geometry to the data; in Sec.~\ref{sec:results} we present our main results and in Sec.~\ref{sec:discussion} we discuss them in the context of the literature; in Sec.~\ref{sec:conclusion} we summarize our results and draw our conclusions.

\section{Data} \label{sec:data}

The main emphasis in this work is given to the modeling of the line emission detected in the first all-sky survey of the \erosita\, data (eRASS1).
However, a comprehensive analysis of the hot CGM of the MW, which is the scientific driver of our research, cannot ignore additional and complementary information retrieved by independent missions or methods. 
Among the data sets presented below, the eRASS1 maps retain the highest statistical power thanks to the orders-of-magnitude larger sample size.

We analyse the data from the first \erosita \, All-Sky Survey of the \erosita{}\_DE consortium. We exploit images of the count rate per pixel generated from the original event files of \erosita \,in different narrow energy bands as presented in Zheng et al., in preparation (please refer here for details on the image production).
The narrow band encompasses the [0.614-0.694] keV range and is named after the most prominent emission line included within the range, namely the O VIII line ($\sim$0.654 keV).
The energy range has been fixed to 80 eV around the line centroid, in order to approximately account for the \erosita\, energy resolution at around the corresponding energies \citep{2021A&A...647A...1P}. The O VIII band map has been created using the eROSITA Science Analysis Software System (eSASS, version 020) and is shown in Fig.~\ref{fig:hmap_O8_n128_id0}.
The eROSITA data have been checked and validated against the RASS data \citep{1997ApJ...485..125S}, finding values consistent with the ones presented here (Zheng et al., submitted). \\
The eRASS1 data feature a highly non-uniform exposure across the sky, producing a direction-dependent sensitivity threshold to point source detection. Removing point-source emission homogeneously (including AGN) and taking the subtraction correctly into account in the modeling of the cosmic X-ray background (CXB) is non-trivial. We thus decide to subtract only the brightest point sources detected in the 0.2-2.3 keV band (Merloni et al. in prep.) above a flux of $\rm 10^{-12}\, erg\, s^{-1}\, cm^{-2}$. This high threshold allows to remove bright source contamination in the maps while keeping the CXB contribute uniform across the sky.
We convert the original FITS maps in \healpix format exploiting the {\small HEALPy} Python library \citep{Zonca2019}. The \healpix maps, defined to provide sky area units of equal surface \citep{2005ApJ...622..759G}, have been used for the analysis, whereas a Zenith Equal Area projection is used throughout this work only for displaying purposes.

In addition, we consider a catalogue of O VII $K\alpha$ absorption column densities retrieved thanks to the higher spectral resolution of the XMM-{\it Newton} Reflection Grating Spectrometer \citep{2007ApJ...669..990B, 2013ApJ...770..118M}.
The column density $N$ of the absorption lines is directly proportional to (the integral of) the density of the medium $N\propto nL$ (whereas emission intensity goes as $\propto n^2L$). For this reason O VII absorption data are particularly relevant to constrain lower density plasma located at large distances from the Sun.

\subsection{eRASS1 data selection}

\begin{figure*}
    \centering
    \includegraphics[width=0.49\textwidth]{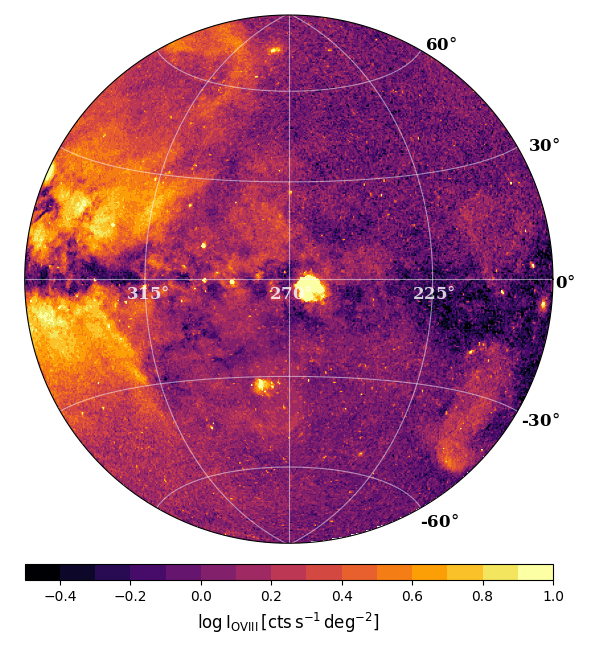}
    \includegraphics[width=0.49\textwidth]{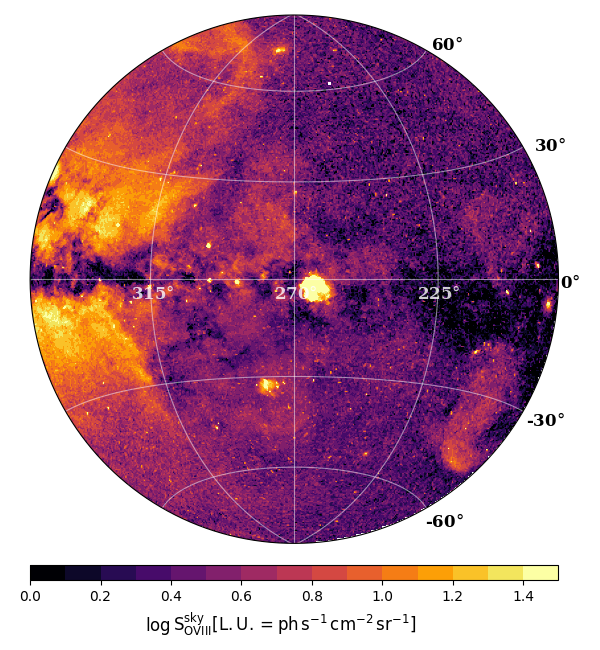}
    \caption{eRASS1 O VIII intensity data used in this work. Data are shown using a Zenith Equal Area projection in the Galactic coordinate reference frame and centered at $(l,b)=(270,0)\, \rm deg$ throughout this work. We note that the left panel includes the contribution from the \erosita{} instrumental background and the sky flux is processed through the effective area of \erosita{}. The right panel instead shows the same map where the instrumental background has been subtracted, leaving the signal from the sky (see eq.~\ref{eq:I_sky}). The sky signal has then been divided by the \erosita{} effective area to obtain the surface brightness of the physical sky signal in line units ($\rm L.U. = ph\, s^{-1}\, cm^{-2}\, sr^{-1}$). }
    \label{fig:hmap_O8_n128_id0}
\end{figure*}
The eRASS1 data cover the Western Galactic sky. We show the O VIII narrow energy band map in Fig.~\ref{fig:hmap_O8_n128_id0}. Several large angular scale features and sources show up in the map (e.g. the \erosita{} bubbles, the Eridanus-Orion superbubble, the Monogem Ring supernova remnant, etc.). These extended sources would bias our analysis if kept in the data. The analysis of the hot CGM of the MW presented in this work requires to exclude the photons coming from these sources. We thus select all pixels within $2\sigma_b$ of Fig.~\ref{fig:hmap_O8_n128_id0} (left panel), where $\sigma_b$ represents the root-mean-square value computed at every latitude $b$ in the ${\rm 220\, deg}<l<{\rm 250\, deg}$ stripe at the same latitude $b$. The longitude range considered is in fact clear from extended foreground sources, as evident from Fig.~\ref{fig:hmap_O8_n128_id0}. 

In addition, we exclude all regions holding total hydrogen column densities $\rm N_H>1.6\times 10^{21}\,cm^{-2}$. Overestimation of the total hydrogen column density may in fact affect lines of sight in which not all of the emission comes from behind the absorption layer. This effect may become more prominent closer to the disk, where column density are also higher. The column density threshold of $\rm N_H>1.6\times 10^{21}\,cm^{-2}$ fixes the bias on the absorbed emission model in the O VIII band to be $<30\%$ for any absorbed source, for assumed $\rm N_H$ values within a factor $+50\%$ of the true value. The selected lines of sight used in our analysis are shown in Fig.~\ref{fig:hmap_valid_n128_id7} in the Appendix. They are mostly found at $|b|>15$ deg. 
Cluster of galaxies from the MCXC catalogue \citep{2011A&A...534A.109P} are also masked up to twice their $R_{500}$ reported value. 
The regions excluded from the resulting mask well approximate either the presence of extended foreground sources or high column density regions (i.e. the Galactic disk). 
The final selected area used for the analysis presented in this work amounts to $\sim 1/3$ of the Western sky (i.e. $\rm \sim6.6k\,deg^2$).

\subsection{Ubiquitous background and foreground components}

The selected data represents what commonly is referred to as the X-ray sky background. The spectrum extracted in the eFEDS region is shown in Fig. 5 of \cite{2022arXiv221003133P}, together with the proposed best-fits of the data. By looking at their model components, the diffuse background in the 0.6-0.7 keV band includes the emission of the hot CGM (blue line), whose 3D structure we aim to analyse. Other components that we aim to isolate are also present, namely the instrumental background (INST, black line) and the CXB (magenta line). In softer energy bands (e.g. O VII, 0.5-0.6 keV) the addition of the Local Hot Bubble emission (LHB, red line) and the potential presence of the Solar Wind Charge Exchange (SWCX, cyan line) complicate the analysis. Despite the O VII is the most prominent CGM emission line, in this work we do not analyze the 0.5-0.6 keV energy range due to the limited knowledge about the detailed morphology of the LHB and SWCX components. Instead, both the LHB and SWCX components are found to only provide a minor contribution in the O VIII band. We then consider the O VIII band as the main driver of our results. The study of O VIII/O VII line ratio is also expected to provide deeper insight into the temperature distribution across the sky, provided that degeneracy between the various components building the O VII emission can be correctly separated. The study of the O VII line however goes beyond the scope of the analysis presented here and will be addressed in a future work.

In addition to a warm-hot medium, a hotter plasma component ($kT\sim 10^7$ K) is introduced in the eFEDS spectral analysis to model excess emission found at $\sim 1$ keV. This component may produce emission also in the O VIII band, although with an intensity comparable to the LHB and SWCX components, amounting to only $1-2 \%$ of the total intensity in this band. Given the very little information available on the shape and properties of this component, as well as the minor contribution in the O VIII band, for simplicity we leave it out from the analysis carried in this work.

We thus describe the CGM intensity as 
\begin{align}
    I_{\rm CGM}(\Vec{s},E)\,e^{ - \sigma(E) N_{\rm H}(\Vec{s}) } &= I^{\rm sky}_{\rm eRASS1}(\Vec{s},E) - I_{\rm add}(\Vec{s},E) \label{eq:I_tot} \\
    I^{\rm sky}_{\rm eRASS1}(\Vec{s},E) &= I_{\rm eRASS1}(\Vec{s},E) - I_{\rm INST}(E) \label{eq:I_sky} \\
    I_{\rm add}(\Vec{s},E) &= I_{\rm CXB}(E)\,e^{ - \sigma(E) N_{\rm H}(\Vec{s}) } + \notag \\
    & + \left[ I_{\rm LHB}(\Vec{s},E) + I_{\rm SWCX}(\Vec{s},E) \right] \label{eq:I_add}
\end{align}
where $\Vec{s}\equiv (l,b)$ is expressed in Galactic coordinates and $E$ is the photon energy. 
The exponential term models the absorption by neutral material in the ISM with $\sigma$ being the absorption cross section of X-ray radiation \citep{1992ApJ...400..699B} assuming photospheric abundances \citep{2003ApJ...591.1220L} and $N_{\rm H}(l,b)$ is the total hydrogen column density along a given line of sight (see Appendix \ref{sec:appendix_A}). The terms in square brackets are considered in the analysis of the O VIII band eRASS1 data but only provide minor contributions to the intensity (see Appendix for further details).

Below we provide descriptions of the components other than the CGM, with the aim to model and subtract them from the eRASS1 intensity, in the different narrow energy ranges.

\subsubsection{Instrumental particle background} \label{subsec:instr}

The spurious contribution of charges produced into the CCD has been computed taking data with the filter wheel in closed position. The resulting particle spectrum has been described and modeled in \cite{2021A&A...647A...1P} and the models for all the telescope modules (TM) have been made available \footnote{https://erosita.mpe.mpg.de/edr/eROSITAObservations/EDRFWC/} with the \erosita \, Early Data Release (EDR) of the Calibration and Performance Verification (CalPV) phase (see also \citealt{Yeung2023}). We built the particle spectrum in units of $\rm cts\, s^{-1}\, deg^{-2} \, keV^{-1}$ by summing the models for the only TMs used to produce our line images (i.e. TM 1, 2, 3, 4 and 6). The spectrum has been then multiplied by the pixel area in deg$^2$ and integrated in the line energy range of interest to obtain the expected count rate. The background has been observed to be significantly constant over the sky, we thus assume single values of $0.122\, \rm cts \, s^{-1}\, \deg^{-2}$.
From the scatter in the TMs data, we estimated a fractional error of 5\% on the instrumental noise component, measured on a deg$^2$ scale.

\subsubsection{Cosmic X-ray background}

We model the cosmic X-ray background as a sum of two broken power-law  \citep{2007PASJ...59S.141S, 2012ApJS..202...14H}, as this model is able to describe the steepening of the summed spectrum of the sources building up the extragalactic background below 2 keV \citep{1993A&A...275....1H}. The first power-law breaks at $E_b = 0.4$ keV and evolves from $\propto E^{-1.9}$ for $E\leq 0.4$ keV to $E^{-1.6}$ for $E\geq 0.4$ keV. At $E>1.2$ keV, after a secon break the function is described by $E^{-1.45}$. The CXB spectrum normalization at 1 keV is 10.9 $\rm ph \, s^{-1} \, cm^{-2} \, sr^{-1} \, keV^{-1}$.
After multiplying by the effective area in the line energy range and integrating over the spectrum, we estimate CXB contributes of 0.29 $\rm ph \, s^{-1} \, deg^{-2}$ in the O VIII band. 
In addition, the CXB is expected to be absorbed by the neutral material in the ISM, as it is generated in the background of the ISM with respect to us. We thus apply the absorption factor $e^{-\sigma(E)N_H(\Vec{s},E)}$ to this component (see Eq.~\ref{eq:I_add}). Due to absorption, the final (additive) CXB contribute to the line intensity is dependent on the line of sight. 
We attribute a fractional error of 4\% to the knowledge of the CXB count rate on a deg$^2$ scale \citep{2003A&A...411..329R}.

\section{Model of the eRASS1 CGM maps} \label{sec:model}

\subsection{Physical models}

By modeling the warm-hot CGM we refer to either an extended spherical halo component or to an oblate or disk-like distribution usually referred to as the "corona". The term corona is used across different fields of astronomy and plasma physics, and is usually linked to the presence of diffuse and highly ionised plasma. For the hot phase of galaxies CGM it has been used in connection with a disk-shape morphology, closer to stellar/ISM scale lengths than to virial radius scales. If the link between a spherical large-scale morphology and the presence of diffuse $\sim$virialized gas is well motivated and supported by other observations (\citealt{2013ApJ...773...92H, 2013ApJ...770..118M} although see also \citealt{2021ApJ...922..121L}), in general, observing a non-spherical disk-like component of the emission does not directly imply a diffuse volume-filling plasma as its source. We thus prefer the geometrical term "disk-like". This semantic difference allows us to separate the direct results of our analysis, which are mostly based on geometrical arguments and to some extent independent on the actual physics, from the physical picture derived after assuming a diffuse plasma as the source of the emission.
The relative importance of the two classes of models (i.e. spherical halo and disk-like), as well as the sources of the emission, motivates us to try different recipes to first describe the geometry of the soft X-ray emission described by the \erosita{} data. 

\subsection{Thermo-chemistry}

We model the gas of the Galactic halo as an isothermal distribution with $T=1.7 \times 10^6$~K ($kT = 0.15$~keV). This assumption on the temperature is supported by the analysis of the \erosita{} spectrum in the eFEDS field \citep{2022arXiv221003133P}, and tightly constrained by the ratio between the line intensities of highly ionized species \citep{2012ApJS..202...14H}. 
The large background region available in the eFEDS field allowed to reach detailed energy calibration of the O VII line in the eRASS1 data, showing a value consistent with being dominated by emission by the recombination line \citep{2022arXiv221003133P}. This corroborates the assumption of a collisionally ionized plasma in thermal equilibrium for the warm-hot component. The G-ratio\footnote{G=(f+i)/r, where f, i and r are respectively the intensities of the forbidden, intercombination and recombination lines \citep{2000A&AS..143..495P}.} and spectral fit using an \apec model also provide consistent and independent temperature estimates of $kT=0.15-0.17$ keV (respectively including or not a SWCX component).
Different studies find a warm-hot temperature component at slightly different temperatures around $kT\simeq 0.2$ keV \citep{2002ApJ...576..188M, 2009PASJ...61..805Y, 2021ApJ...909..164G, 2021ApJ...918...83D, 2020NatAs...4.1072K, 2022arXiv220807863B}, however, the systematic shifts may arise depending on the number of gas phases used to fit of the spectra \citep{2022ApJ...936...72B}.
Previous studies also found a small observed scatter at high Galactic latitudes ($\Delta kT\simeq 0.023$ keV, \citealt{2020NatAs...4.1072K}).
In addition, we test a model with $kT=0.225$ keV and $Z=0.3\, Z_\odot$ to assess potential systematic uncertainties in the fit results.
We note that a constant temperature profile is also expected for any virialized halo following a total mass profile $M(<r) \propto r$, such as the one derived by assuming a Navarro-Frenk-White dark matter profile \citep{1997ApJ...490..493N} in the theoretical framework of $\Lambda$CDM cosmology. Important deviations from the virial temperature may anyway be common depending on the amount of turbulence, bulk motions, magnetic fields and accelerated cosmic rays in one galaxy.
A uniform metal abundance $Z=0.1 \, Z_\odot$ is also assumed for the halo \citep{2022arXiv221003133P}.

\subsubsection{Geometry: spherical $\sim$ virialized halo}
In the extended halo model the density of the hot material is most simply described by a spherical $\beta$ model 
\begin{equation}
    n(r) = n_0 \left[ 1 + \left( \frac{r}{r_0} \right)^2 \right]^{-\frac{3}{2} \beta} \label{eq:beta}
\end{equation}
where $r$ is the distance to the Galactic center, $n_0$ and $r_0$ describe the flattening of the inner profile, while $\beta$ describes the roll off of the density at large radii. In fact, since in practice we masked out most directions in the quarter slab close to the Galactic center ($|l|\geq 270\,\deg$) due to either foreground structures or high absorption, a simpler formula for the $\beta$ model is considered, by taking the limit of the model to large radii. Small radii are in fact probed mostly by directions close to the Galactic Centre, excluded from our analysis. The asymptotic formula also reduces the number of degree of freedom as
\begin{equation}
    n(r >> r_0) = n_0 r_0^{3 \beta} r^{-3\beta} = C \, r^{-3\beta} \label{eq:beta_asympt}
\end{equation}
This treatment allows to ignore the degeneracy between the central parameters and to potentially obtain a more robust estimate for $\beta$ which is key to correctly infer the mass of the baryons. 

\subsubsection{Geometry: oblate disk-like component}

In an alternative scenario the X-ray emission can be produced by a hot corona powered by outflows of hot gas driven by supernovae explosions, surrounding the stellar disc, or by an unresolved population of Galactic sources distributed in and around the disk. Both imply a non-spherical geometry and are expected to mimic the flatter distribution characteristic of the stellar/ISM disc although with potentially different scale length/height. 
To model this kind of flattened density distribution, we can either set independent scale heights along the radial direction (R direction) and perpendicularly to the mid-plane (z direction).

Plasma processed by supernovae expanding from the stellar disc and/or condensating fountains of material falling back to the disk from where it was expelled, both contribute to creating a hot and thick atmosphere around the stellar disk. Hydrodynamic (non-)equilibrium arguments imply a steeper decrease of the density with the distance with respect to the beta model. It makes sense then to model this kind of disk-like extended corona with an exponential function of the radius rather than a power-law model. We describe this oblate/disk-like atmosphere with the following model \citep[e.g.][]{2009ApJ...690..143Y, 2017ApJ...849..105L}.
\begin{equation}
    n(r) = n_0 e^{ -R / R_h } e^{ - |z| / z_h } \label{eq:LB17}
\end{equation}

We note that a similar (i.e. exponential) distribution is expected also in the case that the emission is related to an unresolved population of hot stars in the disk rather than a truly diffuse plasma. In this case, the scale height will be related to the actual distribution of such star population. The topology of the resulting emission can be modelled similarly as a thick disk.
We neglect flattened distributions tilted with respect to the plane of the galaxy for simplicity.

\subsection{Estimated model intensity}

The model in Eq.~\ref{eq:I_tot} is described in Galactic coordinates and thus assumes the observer is located at the position of the Sun, at distance $R_0=8.2$ kpc from the Galactic center. However, as can be seen from Eq.~\ref{eq:beta} to~\ref{eq:LB17}, the models are first defined in the reference frame of the Galactic center. They thus have to be transformed to the Sun reference frame through the following set of equations \citep{2015ApJ...800...14M}:
\begin{align}
    R^2 & = R_0^2 + s^2 \cos^2(b) - 2 R_0 \cos(b) \cos(l) \\
    z^2 & = s^2 \sin^2(b) \\
    r^2 & = R^2 + z^2
\end{align}
where we have called $r$ the distance to the Galactic center and $s$ the distance relative to the Sun.
The change of reference frame is crucial and introduces a direction dependence of the emission morphology even for plasma geometries spherically symmetric around the Galactic Center.

Given our assumption of constant temperature, the density of a plasma component at each point (i.e. a volume's voxel) is converted to an emission profile through a constant emissivity $\epsilon(T)$ depending on the assumed gas temperature. 
We compute the line emissivities from an \apec \citep{2001ApJ...556L..91S} emission model and smooth its spectrum with a gaussian kernel with $\rm FWHM \simeq 80$~eV in order to mimic the energy resolution of \erosita. 
We compute $\epsilon(0.15 \rm keV )= 1.649$ and $\epsilon(0.225 \rm keV )= 2.607$ in units of $10^{-15} (Z/Z_\odot)\, \rm ph \, cm^3 \, s^{-1}$ in the O VIII band (dominated by the O VIII line emissivity).
We then integrate over all points located along one line of sight described by a vector in Galactic coordinates $\Vec{s} = (l,\,b,s)$ at a distance $s$ from the Sun, up to a maximum distance $R_{\rm out}=350$~kpc
\begin{equation}
    I_{\rm CGM}(l,b) = \frac{1}{4 \pi} \int_0^{R_{\rm out}} n^2(\Vec{s}) \epsilon(T)\, d\Vec{s}. \label{eq:I_CGM}
\end{equation}

We note that as long as the $n^2(s)$ profile is steeper than $s^{-2}$, we have that the choice of $R_{\rm out}$ does not affect the integral. For flatter profiles instead, $R_{\rm out}$ can weakly affect the total CGM emission and mass, up to divergence for profiles equal or flatter than $s^{-1}$. The above conditions ($s^{-2}, \, s^{-1}$) are met for $\beta$ models holding $\beta=1/3$ and $\beta=1/6$ respectively. High values of $\beta>1/3$ imply that the bulk of the emission is provided by gas well within $R_{\rm out}$, whereas $\beta<1/6$ profiles will hold the bulk of the mass and the emission at the outer boundary, diverging for $R_{\rm out}\longrightarrow \infty$. We thus consider as non-physical all values $\beta<1/6\simeq 0.17$.
In our analysis we neglect optical depth corrections and assume the gas to be optically thin.\\

To fit the model map to the data, we use a Markov Chain Monte Carlo bayesian algorithm implemented in Python language by the {\tt ultranest} library \citep{2021JOSS....6.3001B}. 
We define our Likelihood $\mathcal{L}$ to be maximized (we actually minimize its logarithm) by the best-fit solution as
\begin{align}
    \ln \mathcal{L}(I_{\rm model} | l,b, \Vec{\theta}) \equiv -\frac{1}{2\nu} \sum_{l,b}  \left( \frac{I_{\rm obs}(l,b) - I_{\rm model}(l,b, \Vec{\theta})}{\sigma_{\rm obs}(l,b)} \right)^2
\end{align}
for any set of parameters $\Vec{\theta}$, where $\sigma_{\rm obs}$ is the uncertainty on the data $I_{\rm obs}$.

\section{Results} \label{sec:results}

The fit results of the models presented in Sec.~\ref{sec:model} are reported in Table 1 and commented with some minimal notes to convey the main messages of each model fit. In this section we provide a more detailed description of the broad classes of models tested: the spherical $\beta$ and the exponential disk-like models.

\begin{table*}
    \centering
    \small
    \begin{tabular}{ l l l l l l l }
        name & profile & parameters & $\chi^2$ & d.o.f. & $\chi^2/$d.o.f. & notes \\
        \hline
        \hline
        spherical $\beta$ & eq.~\ref{eq:beta_asympt} & \multicolumn{1}{p{4cm}}{ $\beta=0.26\pm 0.01$ \newline C$=(10.9\pm 0.6)\times 10^{-3}\, \dagger$ } & 59277 & 31418 & 1.89 & \multicolumn{1}{p{4cm}}{ diverging intensity $(\beta<1/3)$; \newline poor fit at low $|b|$ } \\
        \hline
        disk-like & eq.~\ref{eq:LB17} & \multicolumn{1}{p{4cm}}{ $n_0=(1.7\pm 0.1)\times 10^{-2}\, \rm cm^{-3}$ \newline $R_h = 8.0\pm 0.3$ kpc \newline $z_h=3.3\pm 0.1$ kpc } & 52028 & 31417 & 1.66 & \multicolumn{1}{p{4cm}}{ good fit of the intensity; \newline underestimates $N_{\rm OVII}$ } \\
        \hline
        combined ($\beta\equiv 0.3$) & eq.~\ref{eq:beta_asympt} + \ref{eq:LB17} & \multicolumn{1}{p{4cm}}{ $\beta \equiv 0.3$ \newline C$=(0.9\pm 1.0)\times10^{-3}$ \newline $n_0=(1.7\pm 0.1)\times 10^{-2}\, \rm cm^{-3}$ \newline $R_h = 8.0\pm 0.4$ kpc \newline $z_h=3.3\pm 0.2$ kpc } & 52028 & 31416 & 1.66 & $\sim$flat $\beta$ model is suppressed \\
        \hline
        \bf combined ($\beta\equiv 0.5$) & {} & \multicolumn{1}{p{4cm}}{ \bf $\bf \beta \equiv 0.5$ \newline C$\bf =(4.6\pm 0.1)\times10^{-2}$ \newline $ \rm  \bf n_0=(3.2\pm 0.4)\times 10^{-2}\,cm^{-3}$ \newline $\bf R_h = 6.2\pm 0.4$ kpc \newline $\bf z_h=1.1\pm 0.1$ kpc } & \bf 51805 & \bf 31416 & \bf 1.65 & \multicolumn{1}{p{4cm}}{ \bf $R_h\simeq$ MW stellar-disk radius } \\
        \hline
        combined ($\beta\equiv 0.7$) & {} & \multicolumn{1}{p{4cm}}{ $\beta \equiv 0.7$ \newline C$=(17.6\pm 0.4)\times10^{-2}$ \newline $n_0=(1.6\pm 0.2)\times 10^{-2}\, \rm cm^{-3}$ \newline $R_h = 9.9\pm 0.9$ kpc \newline $z_h=1.6\pm 0.1$ kpc } & 51794 & 31416 & 1.65 & \multicolumn{1}{p{4cm}}{ large central density of \newline the spherical halo } \\
        \hline
        combined+swcx & {} & \multicolumn{1}{p{4cm}}{ $\beta \equiv 0.5$ \newline C$=(3.2\pm 0.1)\times10^{-2}$ \newline $n_0=(6.3\pm 0.8)\times 10^{-2}\, \rm cm^{-3}$ \newline $R_h = 3.9\pm 0.2$ kpc \newline $z_h=0.9\pm 0.1$ kpc } & 51448 & 31416 & 1.64 & \multicolumn{1}{p{4cm}}{ including SWCX model, \newline $kT=0.178$ keV } \\
        \hline
        combined+20\%instr & {} & \multicolumn{1}{p{4cm}}{ $\beta \equiv 0.5$ \newline C$=(4.3\pm 0.1)\times10^{-2}$ \newline $n_0=(5.5\pm 0.6)\times 10^{-2}\, \rm cm^{-3}$ \newline $R_h = 4.6\pm 0.3$ kpc \newline $z_h=0.9\pm 0.1$ kpc } & 51836 & 31416 & 1.65 & \multicolumn{1}{p{4cm}}{ $+20\%$ INST (or CXB) intensity \newline {} } \\
        \hline
        combined+high$\epsilon$ & {} & \multicolumn{1}{p{4cm}}{ $\beta \equiv 0.5$ \newline C$=(1.36\pm 0.03)\times10^{-2}$ \newline $n_0=(9.4\pm 0.9)\times 10^{-3}\, \rm cm^{-3}$ \newline $R_h = 6.2\pm 0.4$ kpc \newline $z_h=1.1\pm 0.1$ kpc } & 51803 & 31416 & 1.65 & $Z=0.3 Z_\odot, \, kT=0.225$ keV \\
        \hline
    \end{tabular} \label{tab:results}
    \caption{ Summary of the best-fit results for the different models. $\rm \dagger\, C\equiv n_0 r_0^{3\beta}\, [cm^{-3}\,kpc^{3\beta}]$ (see eq.~\ref{eq:beta_asympt}) can be interpreted in terms of a central density $\rm n_0\, [cm^{-3}]$ by assuming $r_0\equiv 1$ kpc. }
\end{table*}

\subsection{Combined $\beta$ + disk geometry}

\begin{figure*}
    \includegraphics[height=5.7cm]{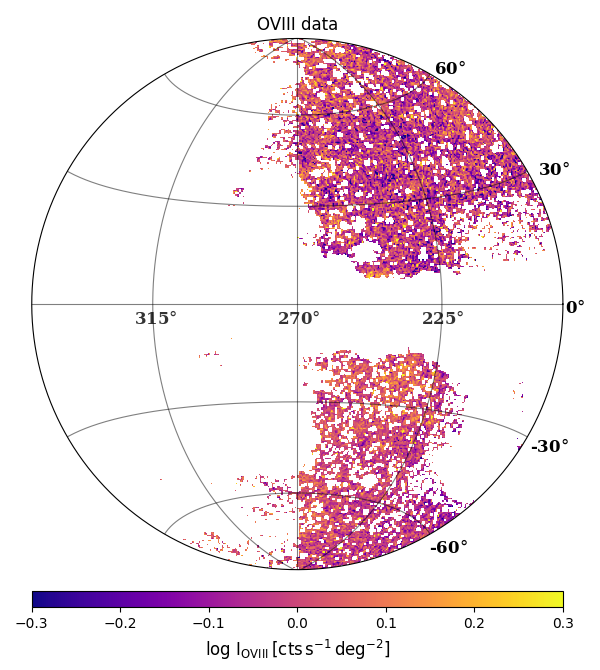}
    \includegraphics[height=5.7cm]{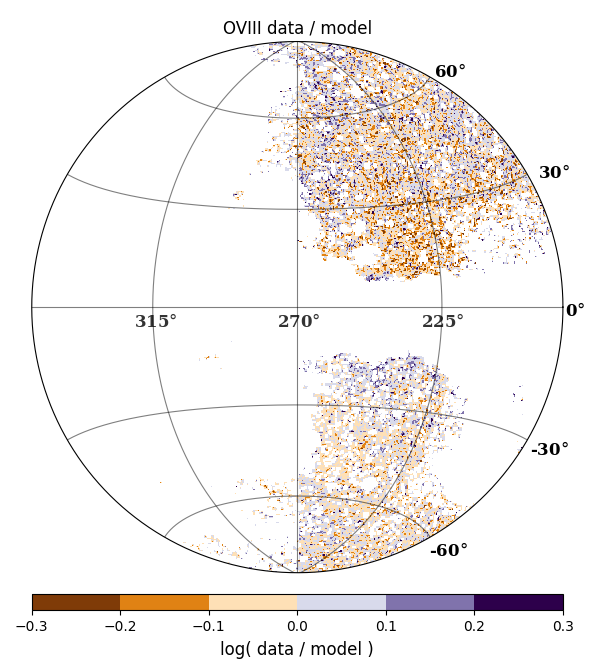}
    \includegraphics[height=5.7cm]{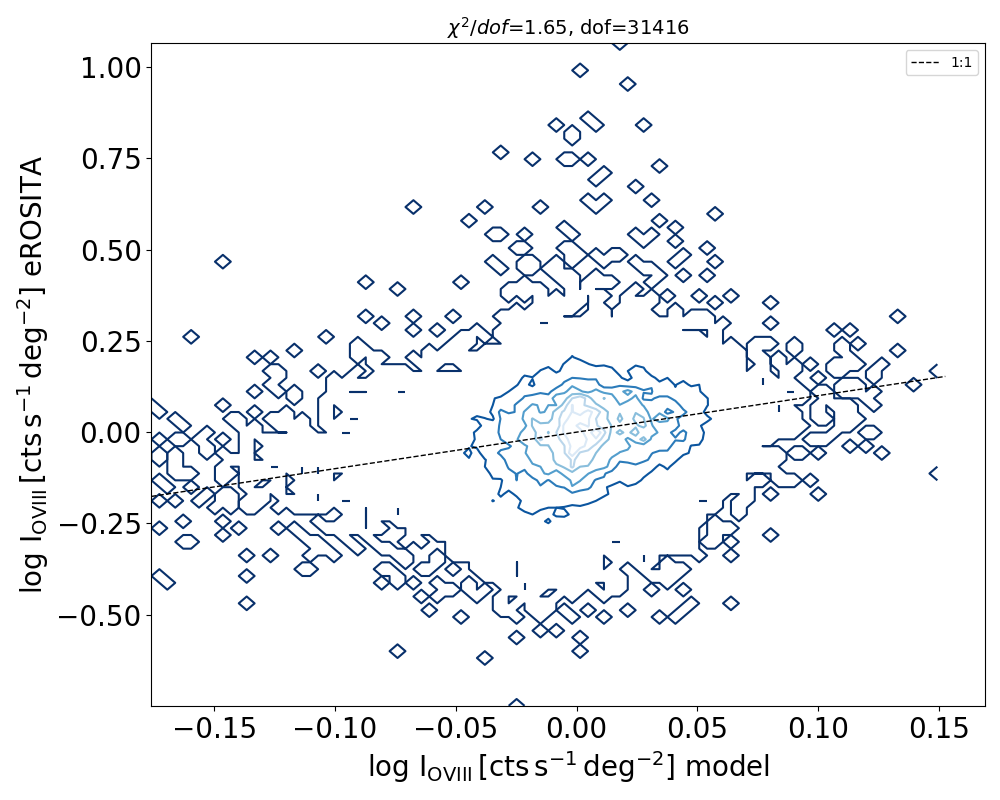}
    \includegraphics[height=5.7cm]{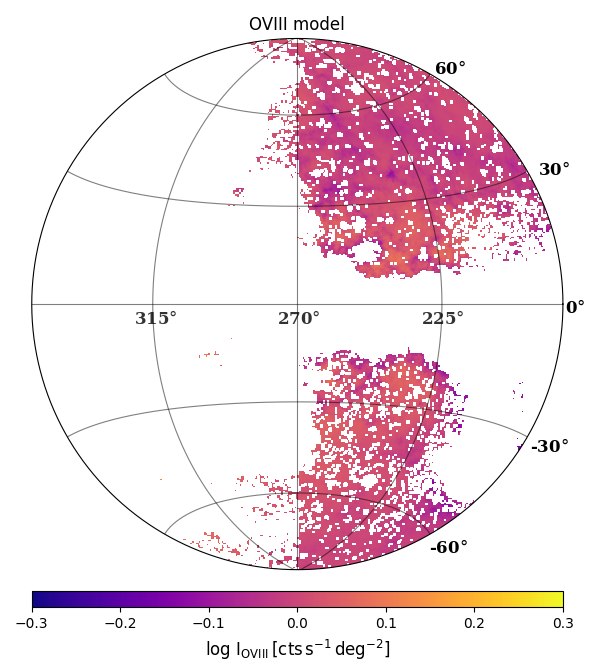}
    \includegraphics[height=5.7cm]{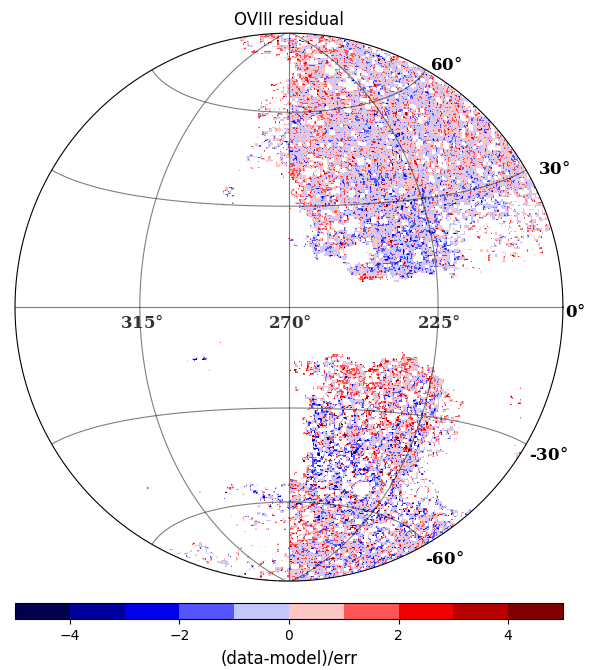}
    \includegraphics[height=5.7cm]{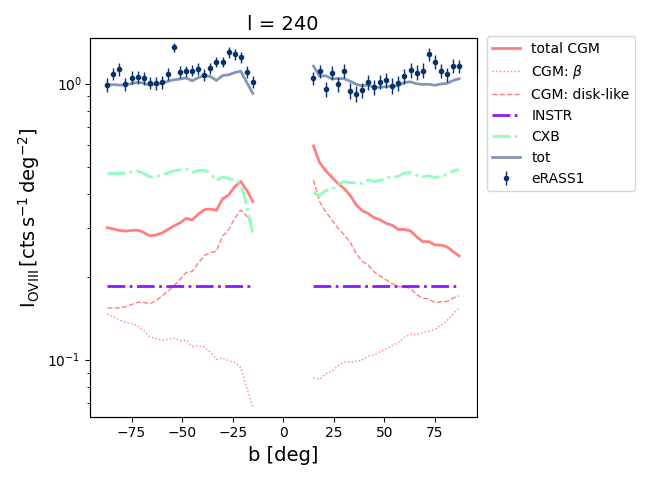}
    \caption{Fit results of the combined ($\beta\equiv 0.5$) density model to the O VIII eRASS1 intensity data. The four maps on the left hand side show the eRASS1 selected data (top left) and the best-fit model intensities (bottom left; including all the background and foreground components, and using the same color scale as the data); the logarithm of the ratio data / model in the range $10^{-0.3}\simeq 0.5$ to $10^{0.3}\simeq 2$ (top center); the residuals (data-model)/error (bottom center). The plots on the right show the data intensities against the predicted best-fit model intensities (top right) and an example latitudinal profile extracted along the $l=240$ deg line (bottom right), highlighting the contribute of the different modelled components.}
    \label{fig:X18O8B05}

\end{figure*}

The current picture in literature describes the X-ray emission and absorption attributed to the hot CGM of the MW as a sum of two components holding a different geometry: a disk-like exponential profile (eq.~\ref{eq:LB17}) accounting for most of the emission and a spherical $\beta$ (eq.~\ref{eq:beta_asympt}) halo producing a minor contribute to the emission, but accounting for most of the absorption of background light.

The O VIII data, the derived best-fit model and its residual are shown in Figures~\ref{fig:X18O8B05}.
We assume gaussian priors for the disk scale-length $R_h=12\pm 4$ kpc and scale-height $z_h=3\pm 0.5$ kpc. The value $R_h=12\pm 4$ kpc has been chosen to include the exponential drop of the HI gas disk of the MW \citep{2008A&A...487..951K, 2017MNRAS.465...76M}. The large standard deviations ($\sigma_R=4$ and $\sigma_z=0.5$) allow a wide range of values, while helping the fit convergence. 
In addition, we fix the slope $\beta$ of the $\beta-$model. To assess the systematic on the fit results introduced by the choice of $\beta$ however, we repeat the fit for different choices of $\beta=0.3,\, 0.5,\, 0.7$.

\begin{figure}
    \centering
    \includegraphics[width=\linewidth]{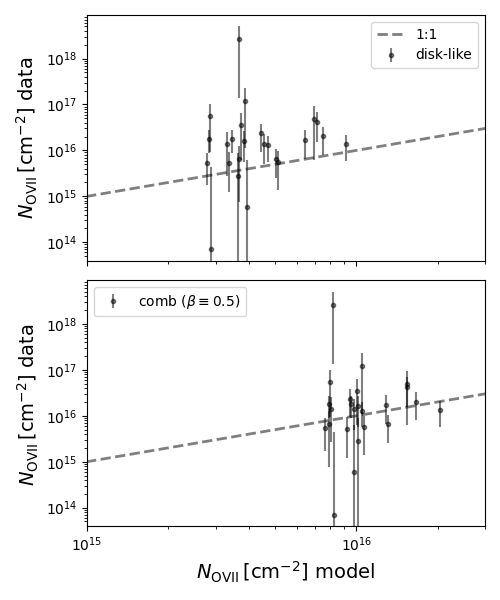}
    \caption{O VII absorption column density data \citep{2013ApJ...770..118M} vs. model. The disk-like model systematically underestimates the absorption column densities. The inclusion of a spherical $\beta$ model allows to account for them. A combination of the two model geometries thus explains both absorption and emission data.}   \label{fig:inspect_MB13}
\end{figure}
The presence of a spherical component allows to account for the column densities derived from O VII absorption lines studies, otherwise systematically underestimated, as shown by Fig.~\ref{fig:inspect_MB13}. The absorption column density of the warm-hot plasma is proportional to the plasma density $\propto n L$ rather than $\propto n^2L$. This makes the length $L$, and thus the plasma length-scale, more relevant with respect to the density in absorption rather than emission data. In fact, a model including only a disk-like component systematically underestimates the observed equivalent widths of $z=0$ O VII lines detected in the spectra of background quasars \citep{2012ApJ...756L...8G, 2020NatAs...4.1072K}.

On the basis of the $\chi^2$ statistic all the realization of the combined model with different $\beta$ are similarly good, with a very small preference ($\Delta\chi^2\simeq -357$ over $\sim 31416$ d.o.f.) for the value $\beta=0.5$ with the inclusion of a SWCX model.
Given this degeneracy, in the following we will consider the combined ($\beta\equiv 0.5$) model as reference for comparison with other results presented in the literature. 
Although we stress that an actual fit of the halo $\beta$ parameter is currently prevented by X-ray intensity data, $\beta\equiv 0.5$ is considered with respect to other solutions also on the basis of theoretical arguments\footnote{in the self-gravitating isothermal sphere (virial) model $\rm \beta\equiv \mu m_p \sigma_{gal}^2 / (kT_{gas}) = 0.5$ is the square of the galaxy-to-gas velocity dispersion ratio (\citealt{1988S&T....76Q.639S, 1962AJ.....67..471K}, although see also \citealt{2022arXiv220609925L}).}. 
Nevertheless, we will take into account the systematic uncertainty introduced by the choice of $\beta$ on other derived quantities (e.g. MW baryonic mass $M_b$ and fraction $f_b$, see Sec.~\ref{sec:discussion}).

\subsection{Oblate disk-like model}

\begin{figure*}
    \includegraphics[height=4.2cm]{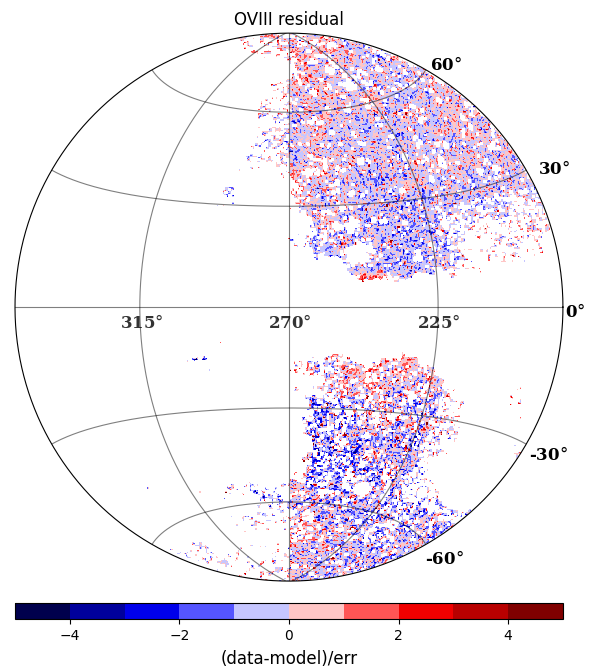}
    \includegraphics[height=4.2cm]{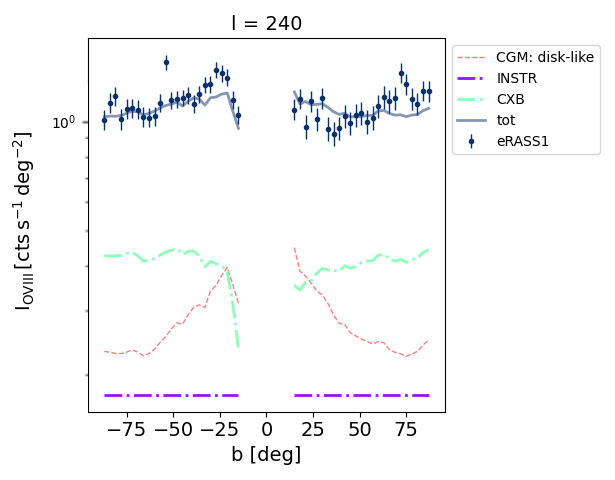}
    \includegraphics[height=4.2cm]{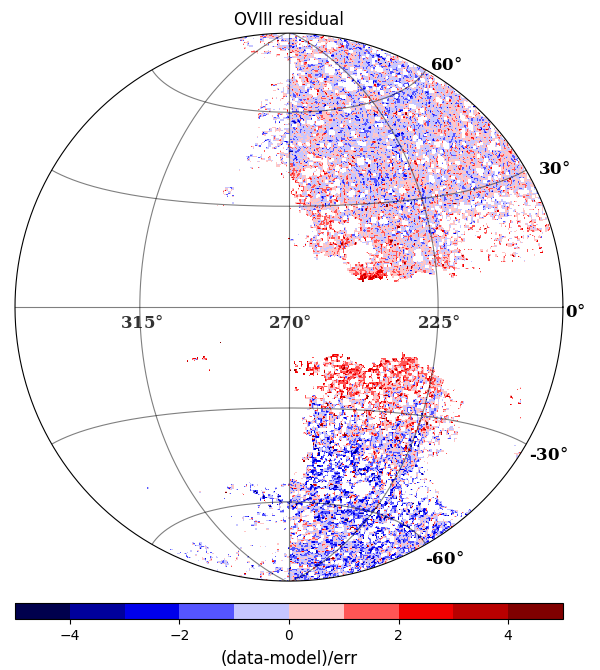}
    \includegraphics[height=4.2cm]{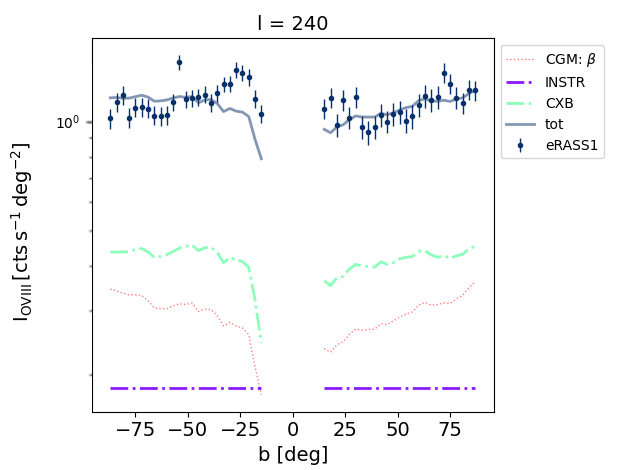}
    \caption{Results of the disk-like (left) and spherical $\beta$ (right) density model fits to the O VIII eRASS1 intensity data. All the plot details are the same as Fig.~\ref{fig:X18O8B05}}
    \label{fig:X9O8}
\end{figure*}

From the results of the combined ($\beta\equiv 0.5$) model, it follows that the majority of the CGM X-ray emission is produced by the disk-like component (see the lower right panel in Fig.~\ref{fig:X18O8B05}). It is thus instructive to see how the fit is affected when using only this disk-like component. This case has the advantage of simplifying the model description, at the expenses of not accounting for the observed O VII absorption.

When looking at the fit residuals (Fig.~\ref{fig:X9O8}) and $\chi^2$ statistic (Table~\ref{tab:results}), the disk-like model overall does not perform significantly worse than the combined ($\beta\equiv 0.5$) model and provides a similar scale length $R_h=8.0\pm 0.3$ kpc. However, the scale height $z_h$ significantly increases to $z_h=3.3\pm 0.1$ kpc.
This is explained by the portion of the emission previously attributed to the halo (increasingly important at high $|b|$, and absent in the disk-like model) now having to be accounted for in the disk-like component alone, in turn increasing the scale height (cfr. the profiles in Fig.~\ref{fig:X18O8B05} and Fig.~\ref{fig:X9O8}).
Based on the only \erosita{} data we can not disfavour a scale height $z_h=3.3$ kpc with respect to 1.1 kpc. However, this can be done by comparing the expected amount of O VII absorption accounted for by the different models. As pointed out above, the disk-like model alone systematically underestimates O VII absorption data.\\

\subsection{Spherical $\beta$ model}

For completeness we also test our data with a fit of a spherical $\beta$  model alone.
Our data selection leaves out most directions at $l>270$ deg. This severely reduces the constraining power on the combination of central density $n_0$ and scale $r_0$ in a spherically symmetric geometry of the plasma. Without the lines of sight towards the Galactic Center these two parameter are in fact highly degenerate with each other. We thus fit the spherical $\beta$ model using its $r>>1$ approximation presented in eq.~\ref{eq:beta_asympt}. 
Looking at the right panels in Fig.~\ref{fig:X9O8}, the $\beta$ model ($\rm \chi^2/d.o.f.=1.89$) can account for the sky average count rate, as the ratio data/model (upper central panel) lays within the 0.5-2 range across most of the sky. However, from the data we observe a systematic increase of the count rates with decreasing $|b|$ at all longitudes that can not be accounted for by a spherical geometry. The same trend is made more evident by looking at the residual (data-model)/err map (Fig.~\ref{fig:X9O8}). 
This evidence alone suggests the presence of an oblate disk-like component other than the INST (flat) and CXB (increasing with $|b|$) as a significant component of the X-ray intensity from the MW (as independently suggested by \citealt{2009ApJ...690..143Y, 2017ApJS..233...20L, 2018ApJ...862...34N} and \citealt{2020NatAs...4.1072K} using {\it Chandra}, XMM-{\it Newton}, Suzaku and HaloSat data respectively). The analysis of the eRASS1 data now provides an unambiguous signal for this oblate component, which is detected with a very high significance. We will discuss the possible nature of this emission in Section~\ref{sec:discussion}.

The very flat best-fit slope of the density distribution of the spherical halo $\beta=0.23$ is likely biased low in order to accommodate the lower rates at high latitudes and the higher rates at lower latitudes together. Such a flat slope would also result in a non physical diverging emission for $r\longrightarrow \infty$. \\
Previous studies focusing on only very high latitude regions found steeper slopes consistent with $\beta=0.5$ \citep{2017ApJ...849..105L}. By fixing $\beta\equiv0.5$, the fit worsens ($\rm \chi^2/d.o.f.=1.97$, not shown), showing even larger residuals at low latitudes and the central normalization increases by an order of magnitude probably to maintain a similar average density (i.e. intensity) value across the volume.
Overall, the spherical $\beta$ model alone reproduces the average intensity of the sky but poorly adapts to the morphology of the eRASS1 O VIII band image.

\subsection{Systematic uncertainties}

We tested the combined models for different choices of $\beta\equiv 0.3,\,0.5,\, 0.7$. The best-fit parameters of the combined models still show some degeneracy with $\beta$. 
The central density of the $\beta$ component C and the scale-length of the disk-like component $R_h$ increase with $\beta$, while the central density of the disk-like component $n_0$ decreases with $\beta$. The scale-height of the disk does not show a clear trend, and is found in the range $z_h\sim 1-3$ kpc. \\
We find a minor but not significant preference for the $\beta\equiv 0.5$ realization of the combined model including the SWCX, based on the $\chi^2$ statistic. 
In the $\beta \equiv 0.3$ combined fit, the $\beta$-model is being suppressed, while the disk-like component dominates the model intensity, showing the same parameters as in the disk-like CGM model. This is probably due to the fact that a $\beta\equiv 0.3$ profile looks flatter across the sky than for higher $\beta$. This brings the dominating trend with $|b|$ in the data to be mainly fitted by means of the other component (i.e. the disk-like), which in turns leaves only a little residual intensity available for the fit of the $\beta$ model. 
The $\beta \equiv 0.7$ fit instead shows a large central normalization C due to the mentioned degeneracy between the various parameters (see Fig.~\ref{fig:X18O8B05_n128_id7_corner} in Appendix). The scale length $R_h$ and height $z_h$ of the disk-like component also increase, due to the density left unaccounted by the steeper roll-off ($\beta=0.7>0.5$).

Three additional models (combined+swcx, combined+20\%instr and combined+high$\epsilon$) are tested to assess the systematic errors introduced respectively by: the introduction of a (minor) SWCX emission component; a potential systematic underestimation of the instrumental noise or CXB component at soft energies; the choice of temperature of the plasma component and the spectral modeling of the soft X-ray emission. \\
In the combined+swcx model we introduce a characterization of the SWCX component in the O VIII band of the eRASS1 data (Dennerl+ in prep., see also Appendix~\ref{appendix:swcx}). 
We compared the estimated CGM flux with an XMM-{\it Newton} measurement obtained after subtraction of SWCX \citep{2007A&A...475..901K}. We consider the latter measurement among the most detailed measurements of the MW CGM component as it relies on a careful model of the SW Parker spiral in space and time, as well as on high spectral resolution data obtained by the Reflection Grating Spectrometer onboard XMM. For the only available field in the Western sky (i.e. the Marano field: $l,b=269.8,\, -51.7$ deg), the authors report $\rm F_{OVIII,cgm}=1.41\pm0.49$ L.U. From the analysis of the eRASS1 data with the inclusion of our SWCX model, we find a consistent value of $\rm F_{OVIII,cgm}^{eRASS1}=1.59\pm0.65$ L.U.

After introducing the SWCX component in the modeling of the eFEDS spectrum the temperature of the CGM component increase from $kT=0.15$ keV to $kT=0.17$ keV \citep{2022arXiv221003133P}. We introduce this change accordingly in the combined+swcx model. 
We estimate differences of $-30,\, +96,\, -37,\, -18\%$ respectively for the $C,\, n_0,\, R_h$ and $z_h$ parameters of combined+swcx with respect to the combined ($\beta\equiv 0.5$) model. The morphological parameters $R_h$ and $z_h$ change mostly due to a non-uniform morphology of the (faint) SWCX component. The shrinked scale-length $R_h$ then requires the normalizations $C$ and $n_0$ to adapt, while accounting in general for the $+16\%$ increase in temperature $kT$ (and in turn emissivity). \\
Similar trends are confirmed when looking at the systematic introduced by a potential underestimation of either the INST or the CXB component (or the sum of them, combined+20\%instr model). Given that these are very bright components in the O VIII band, we want to assess how sensitive are our results with respect to the modeling of these components.
We estimate -6, +19, -26 and -18\% differences with respect to the combined ($\beta\equiv 0.5$) model. The systematic shift is comparable or smaller on all parameters, with respect to the one caused by the inclusion of the SWCX component.\\
In addition, different temperature $kT=0.225$ keV and metallicity $Z=0.3\, Z_\odot$ assumptions on the CGM modeling are tested through the combined+high$\epsilon$ model. Temperature and metallicity both act in practice only on the emissivity $\epsilon(T,Z)$ of the \apec model, that in turn do not affect the fit of morphological parameters ($R_h$ and $z_h$ do not change with respect to the combined ($\beta\equiv 0.5$) model). Both normalization factors $C$ and $n_0$ instead decrease by $\sim 70\%$ compensating for the boosted emissivity $\epsilon_{\rm OVIII}$ due to the higher temperature and Oxygen abundance.\\

The strong dependence of normalization parameters on temperature and emissivity (both fixed quantities in our model) on the one hand stresses even more the large uncertainty affecting the baryon budget encompassed by the spherical halo component (see Sec.~\ref{subsec:mass_profile}), while on the other hand strengthen the evidence for the presence of a disk-like component, as it relies mostly on the morphology of the data, rather than on assumptions on the physical properties of the plasma. This remains true also for a scenario in which (part of) the plasma is at a different temperature than the one assumed in this work, or out of equilibrium. Closeby emission can in principle be produced as well by outflowing hot gas out of thermal equilibrium. However, (lack of) equilibrium mainly affects the gas emissivity, which in turn plays a role only in the determination of the density normalization rather than on the fitted geometry. The $R_h$ and $z_h$ parameters are thus relatively solid against the assumption of a collisionally ionised plasma in thermal equilibrium, unless the distribution of temperature and equilibrium phase changes significantly across the sky.

\section{Discussion} \label{sec:discussion}

In this section we explore some further implications of our fit results, including the effects produced by our assumptions.

\subsection{The disk-like component is everywhere brighter than the spherical halo}

\begin{figure}
    \centering
    \includegraphics[width=\linewidth]{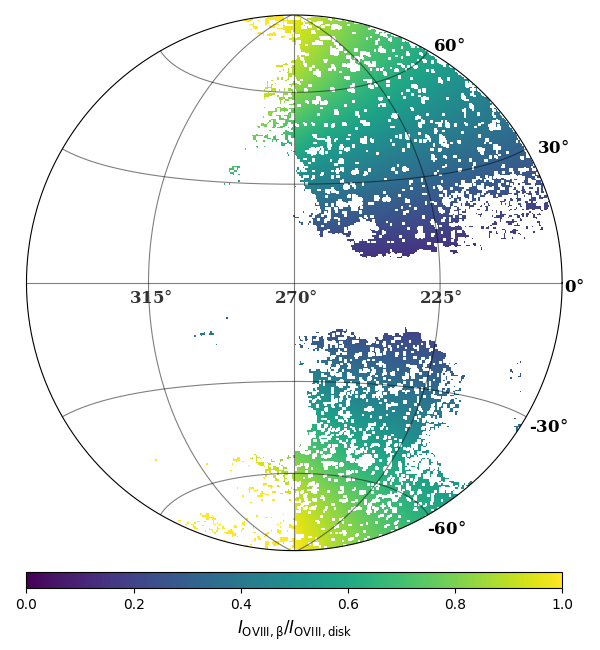}
    \caption{Intensity ratio between the spherical $\beta$ halo and disk-like components in the O VIII band, for the combined ($\beta\equiv 0.5$) model. The disk-like component is everywhere brighter. The two components are about equal only at the Galactic poles.}
    \label{fig:hmap_O8_compRatio_ZEA}
\end{figure}
What is the fraction of the observed X-ray intensity coming from the extended hot halo component? This important question has so far been left unanswered, or only partially answered, due to the limited X-ray intensity data structures available before the advent of \erosita{}. By simulating and re-projecting the expected emission from the components of the combined ($\beta\equiv 0.5$) model (the overall model intensity is shown in the lower left panel of Fig.~\ref{fig:X18O8B05}), we show in Fig.~\ref{fig:hmap_O8_compRatio_ZEA} the ratio between the projected O VIII intensity of the halo ($\beta$) and the disk-like component in the combined ($\beta\equiv 0.5$) model. The ratio is mostly dependent on Galactic latitude $|b|$, with a slight dependence over longitude $l$ too. The ratio is smaller than 1 in all directions, reaching values as low as $10\%$ at $|b|\sim 20$ deg. The ratio confirms that the oblate disk-like component provides most of the counts attributed to the CGM in the O VIII band in all directions. Only at high $|b|>80$ deg the spherical and oblate components result in about the same emission. Previous analyses of high latitude soft-X background data \citep{2020NatAs...4.1072K} have already pointed out the improvement in the fit results when including a spherical and a disk-like component rather than a single one. Our result confirm that this procedure is necessary, as the model components hold in general comparable levels of emission, whereas the spherical component only provides a minor contribution at low latitudes.

We stress that the situation described by Fig.~\ref{fig:hmap_O8_compRatio_ZEA}, as most of the results presented in this work, remains true regardless of the nature of the disk-like component. In fact, a population of unresolved sources with a (thick) disk-like geometry, as well as a truly diffuse plasma embracing the stellar-disk or even emission arising from the interstellar medium, can all be explained by a disk-like geometry component of the emission.

\subsection{Most of the plasma emission is produced within a few kpc} \label{subsec:close_emission}

\begin{figure*}
    \centering
    \includegraphics[width=\linewidth]{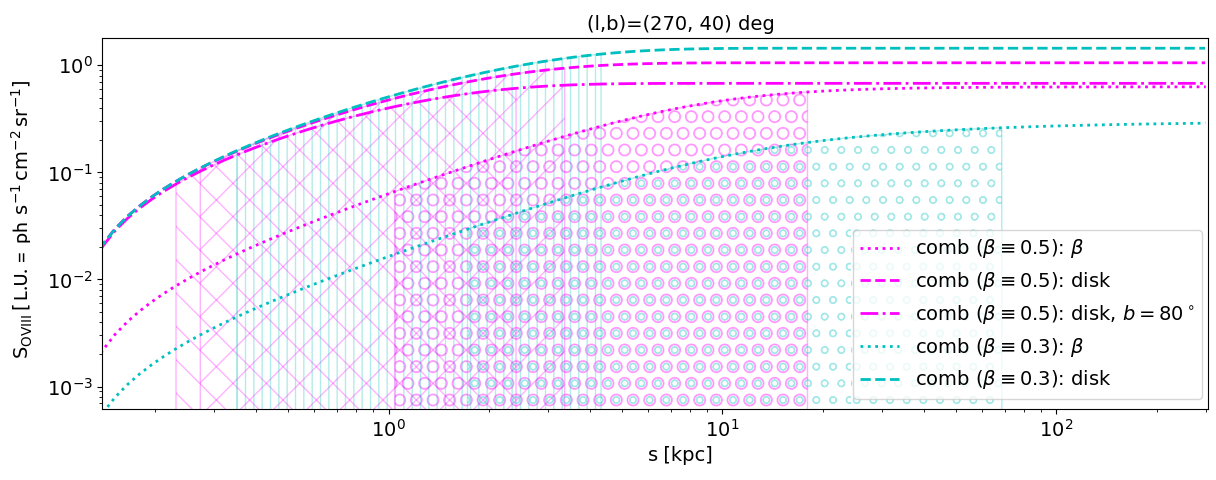}
    \caption{Cumulative surface brightness as a function of the distance from the Sun $s$. For each model, the dotted and dashed lines show the spherical and disk-like components respectively. The hatched areas encompass the $10-90\%$ percentile of the surface brightness distribution.}
    \label{fig:plot_I_vs_r}
\end{figure*}
From the combined ($\beta\equiv 0.5$) model, we predict where most of the observed emission is produced. In Fig.~\ref{fig:plot_I_vs_r} we plot the cumulative surface brightness of the spherical $\beta$ and disk-like components of the combined ($\beta\equiv 0.5$) model as a function of the distance from the Sun $s$. The height of a line at a given distance $s$ tell us in practice how much emission has been produced within that distance by that component of the model.
Given the general non-spherical symmetry of the projected emission, we compute the profiles for $(l,b)=(270,40)$ deg. A line of sight at $b=80$ deg is also shown for the combined ($\beta\equiv 0.5$) model for comparison (dot-dashed line).
The hatched areas encompass the $10-90\%$ percentile of the surface brightness distribution of each component and model. We plot different models to assess the systematic errors. 

We first highlight how the choice of assumed temperature $kT$ and metallicity $Z$ affects the density normalizations but leave the emission profile unchanged in practice. In fact, for simplicity we do not show the combined+high$\epsilon$ model curves in Fig.~\ref{fig:plot_I_vs_r}, as they would overlay entirely with the magenta lines. We already pointed out this feature of the models. Differently, a change in $\beta$ affects the emission profile, mainly distributing the emission in the spherical component over a broader (narrower) and larger (closer) range of distances for lower (higher) $\beta$, while only slightly affecting the disk-like component. In general, the disk-like component always cumulatively produce a larger amount of emission and has a faster increase with $s$. Most of the emission is in fact produced between 0.2 and 5 kpc from us (median $\sim 1$ kpc), including all systematic uncertainties. 

We note that the lower bound of this range is potentially in conflict with our assumption of the absorption happening all in front of the X-ray emission with respect to us, as clouds at latitudes $|b|\sim 20$ deg above the Galactic plane are found up to distances of 1 kpc \citep{2019A&A...625A.135L}. At the lower latitudes, breaking the assumption of a single foreground absorbing layer of colder gas may explain (part of) the positive residual still visible (e.g. Fig.~\ref{fig:X18O8B05}, \ref{fig:X9O8}). Loosening the assumption would unfortunately require to introduce additional degrees of freedom, highly increasing the degeneracy between them. Although we choose not to increase the complexity in our analysis, we point out that only a minor part of the emission in our closest proximity (0.2-0.3 kpc) may be strongly affected by co-spatial absorption, as farther clouds at $|b|>20$ deg only provide a minor contribute to the column density \citep{2019A&A...625A.135L}.

\subsection{The spherical halo holds most of the mass, but its precise budget remains highly uncertain} \label{subsec:mass_profile}

\begin{figure}
    \centering
    \includegraphics[width=\linewidth]{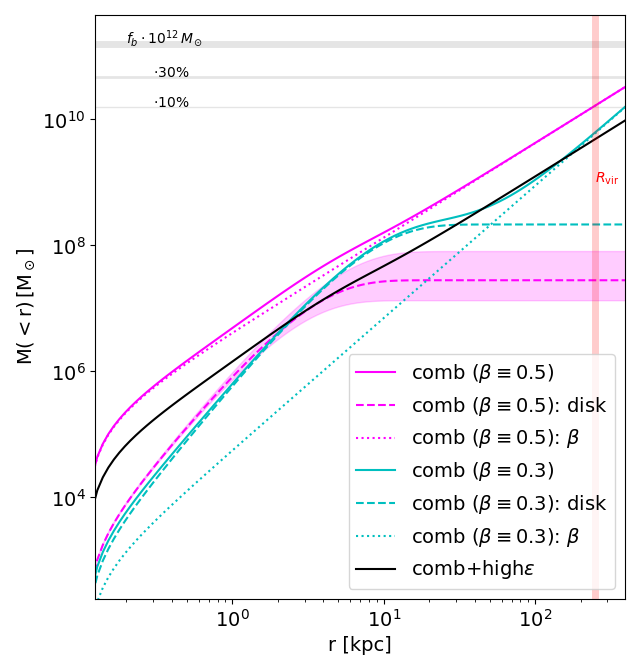}
    \caption{Cumulative mass of the hot gas as a function of the distance from the Galactic Center $r$. For the combined ($\beta\equiv 0.5$) model the solid and dashed lines show the spherical and disk-like components respectively. The horizontal gray lines show 10, 30 and 100\% of $f_b\cdot 10^{12} M_\odot$ where $f_b\equiv \Sigma_b/\Sigma_m=0.071$ is the cosmic baryon fraction.}
    \label{fig:plot_M_vs_r}
\end{figure}
Where is most of the hot gas mass? Given our density models and their components, we integrate over the volume and obtain the mass profiles shown in Fig.~\ref{fig:plot_M_vs_r}. We note that this time we integrate and plot the profile with respect to the distance from the Galactic Center rather than the position of the Sun. Although the disk-like component produces most of the emission, as we have seen above, the component holding most of the mass is the spherical $\beta$ model. This is due to the slower roll-off of the profile with respect to the exponential of the disk. 
The density (and pressure) ratio between the spherical halo and the oblate components increases with distance from the Galactic Center $r$. At large radii, the volume becomes very large $V\propto r^3$, thus collecting most of the mass.
The mass profile becomes steeper for lower (i.e. flatter) $\beta$ profiles, as the density decreases slower for lower $\beta$. Contrary to the surface brightness profiles, in this case the assumptions on temperature $kT$ and metallicity $Z$ of the plasma contribute largely to the systematic offset of the mass profile, as they mainly affect the normalization $C$ and in turn the overall mass. At about the virial radius $R_{\rm vir}\simeq 250$ kpc, depending on the choice of $\beta,\, kT$ and $Z$, the overall mass of hot gas ranges between some $\times 10^9$ to some $10^{10}\,M_\odot$. Thus, the fraction of baryons present in the MW potential well can not be compared to the cosmic fraction $f_b$ to almost an order of magnitude uncertainty. Provided that it makes sense to expect a baryon fraction $f_b$ within the virial radius of one galaxy to be similar to the cosmic value \citep{2021ApJ...922..121L}, and given the very large systematic uncertainty affecting the hot gas mass models, we can only qualitatively infer some fraction of the baryons being seemingly missing from the MW expected budget.

\subsection{The disk-like emission is consistent with the projected mass distribution of the stellar disk}

\begin{figure}
    \centering
    \includegraphics[width=0.49\linewidth]{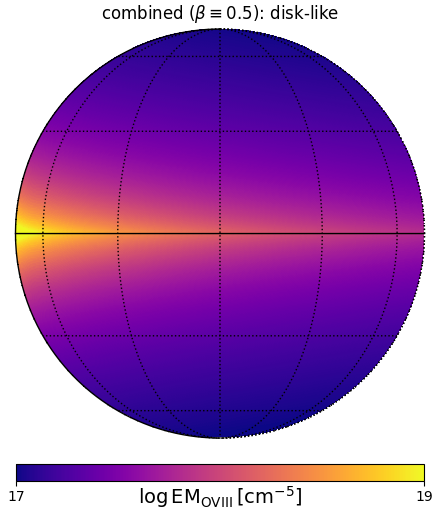}
    \includegraphics[width=0.49\linewidth]{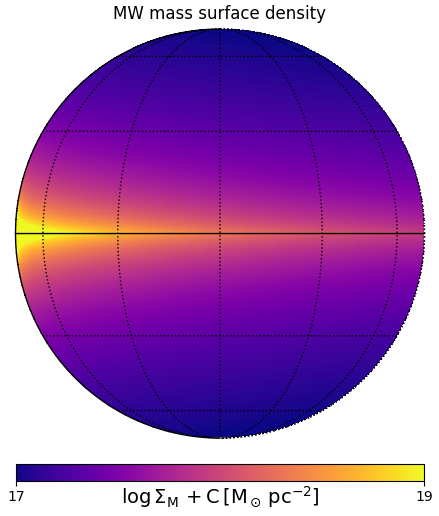}
    \caption{Comparison between the projected morphologies of the combined ($\beta\equiv 0.5$) model EM (left panel) and the MW mass surface density (right panel) profiles. The mass surface density $\Sigma_M$ has been arbitrarily renormalized to show the same range of (log) values [17,19].}
    \label{fig:modelDisk_O8_X18O8B05_n128_id7_EM}
    
    \includegraphics[width=0.7\linewidth]{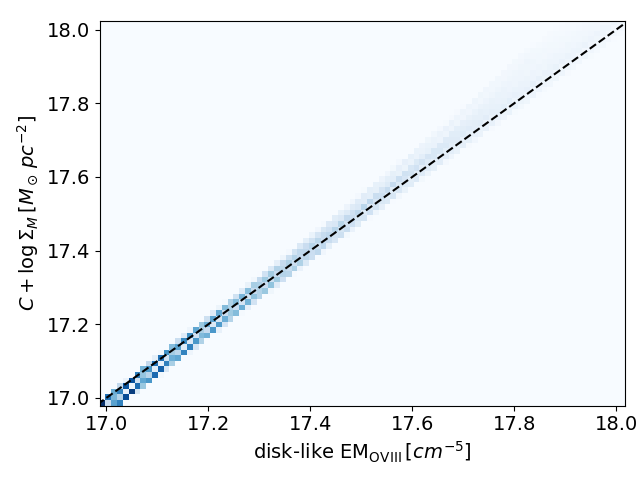}
    \caption{Scatter plot of the quantities shown in Fig.~\ref{fig:modelDisk_O8_X18O8B05_n128_id7_EM}. The dashed black line shows the 1:1 relation between them.}
    \label{fig:plot_MassSurfDens_vs_modDiskO8_n128}
\end{figure}
In this work we have confirmed the existence of a component of the soft-X-ray background emission holding a disk-like geometry (to the 0th order). A straightforward question arises, whether this component can be potentially linked to a population of sources in the MW stellar disk or instead, it truly arises from a diffuse plasma component. Provided that a diffuse disk-like plasma component would be anyway related to stellar populations simply through the shared MW gravitational potential and the origin of the heated plasma, the question can be recast as whether the oblate component is the emission of an unresolved X-ray emitting stellar population or of a diffuse plasma. In fact, as anticipated above, the geometry alone of this component does not allow to exclude one or the other hypothesis. 

The hypothesis of an unresolved M dwarf stellar population producing part of the soft X-ray flux has already been suggested as an explanation for the $\sim 0.7$ keV component, but considered as unlikely due to the smaller scale height attributed to the M-dwarf population \citep{2009PASJ...61S.115M, 2009PASJ...61..805Y}. However, the scale height for the M dwarfs may be larger than what previously assumed, as new models of the mass distribution of the MW disc seems to suggest \citep{2017MNRAS.465...76M}. In addition, M-dwarfs may have an emission component at temperature as low as $kT=0.1-0.2$ keV \citep{2022A&A...661A..29M}, thus also producing O VIII. 
We compute the MW mass surface density profile following \citep{2017MNRAS.465...76M}, projected it at the Sun position\footnote{The complete mass model for the MW must include components as the Nuclear Star Cluster, the Nuclear Stellar Disk, the Galactic bar and the Galactic disk. Since our work selects in practice longitudes $180<l<300$ deg, only the disk component is relevant for our purposes, as all the others only affect the inner $|l|<30$ deg.}. 
The result is shown in the right panel of Fig.~\ref{fig:modelDisk_O8_X18O8B05_n128_id7_EM}, next to the O VIII emission measure (EM) computed for the combined ($\beta\equiv 0.5$) model (left panel). The profiles show a remarkably similar morphology. This is also highlighted by the scatter plot in Fig.~\ref{fig:plot_MassSurfDens_vs_modDiskO8_n128}, showing the two quantities/maps one versus the other. It is very interesting how, despite showing different and completely independent quantities, each derived by completely independent data, the points about follow the 1:1 relation, with some scatter. At high EM values (and high $\Sigma_M$, i.e. close to the Galactic plane), the relation shows some deviation suggesting a slightly different trend. Indeed, the trend is mainly determined by the scale height of the combined ($\beta\equiv 0.5$) model ($z_h=1.1\pm 0.1$ kpc) with respect to the value used to compute the mass profile \citep[$z_h=0.9$ kpc][]{2017MNRAS.465...76M}. Considering the systematic uncertainty on our result, the profiles are consistent with each other.

We note that the normalizations in Fig.~\ref{fig:plot_MassSurfDens_vs_modDiskO8_n128} are clearly not directly comparable, as they originate from different quantities.
Furthermore, the median of the compared models is set equal by definition. The shift $c$ between the EM$_{\rm OVIII}$ and $\Sigma_M$ normalizations may or may not contain information on the luminosity of the stellar population potentially contributing to the X-ray background emission.

Although the geometrical similarity between the X-ray emission of the disk-like component and the projected mass profile looks interesting, we stress that the mass and the combined ($\beta\equiv 0.5$) models compared in Fig.~\ref{fig:modelDisk_O8_X18O8B05_n128_id7_EM} and \ref{fig:plot_MassSurfDens_vs_modDiskO8_n128} are the result of a particular choice among the possible models that we showed being still affected by some systematic uncertainties. In addition, using a geometry-independent approach \cite{2018ApJ...862...34N} estimated the M dwarf unresolved emission from the faint end of their $\log N - \log S$ and found it to account for less than $20\%$ of the CGM flux in the soft X-ray. EM and $\Sigma_M$ may thus eventually differ significantly. \\
The EM of the warm-hot CGM component has also been reported to scale linearly with the EM of the hot component as $\rm EM_{warm-hot} \sim 10.8 \times EM_{hot}$ across high (absolute) latitude regions, after their detection by HaloSat \citep{2022ApJ...936...72B}. 
The authors interpret the EM relation as disfavouring the stellar coronae being responsible for the emission of the hot phase. Their interpretation is based on the assumption that the warm-hot phase is produced by diffuse plasma. By giving up this assumption, the emission of both the warm-hot and the hot phases may be contributed by stellar coronae emission (in part or entirely).
However, both the warm-hot and hot phases have been detected through absorption lines \citep{2019ApJ...887..257D, 2019ApJ...882L..23D}. The detected column densities can not be explained by the small cross section offered by the stellar coronae, while they are easier to accommodate assuming a truly diffuse nature of the plasma phases. While the absorption by the warm-hot component could be accounted for by the spherical halo, the hot phase associated with the disk-like component seems to rule out the unresolved population scenario as its main cause.
Our point here is that the potential connection between the soft-X ray emission and an unresolved stellar population, while being disfavoured by different probes, is not rejected by our simple comparison. In future, a dedicated and more quantitative investigation of the contribute of stellar coronae to the soft X-ray emission, may be worth on the light of new mass models for the MW and the amount and quality of the \erosita{} data. 

However, as already pointed out, the same evidence of a similar scale height between the X-ray disk-like component and the MW mass distribution can be explained by an X-ray emitting gas whose dynamic is governed by the same gravitational potential followed by the stellar thick disk, producing similar scale heights. In this picture, the hot atmosphere is also expected to be stationary to first approximation, as a single episode of energy injection (e.g. an outflow from an active star forming region) is not necessarily expected to correlate with the thick disk height. In the next session, we try to assess if such a hot gaseous disk-like component can be supported by the current stellar activity in the MW disk.

\subsection{The local star formation rate can sustain the disk-like component}

What is the thermal luminosity implied by the emitting gas and how does it compare with the luminosity implied by star formation (i.e. supernovae explosions)?
In Sec.~\ref{subsec:close_emission} we derived that most of the observed emission comes from within few kpc from the Sun. We thus compute the X-ray luminosity in the solar neighbourhood. We consider a cylinder of radius $\rm \Delta R=3$ kpc centered on the Sun position and extending for $\rm \Delta z=3$ kpc above and below the MW mid-plane. The mean density value weighted for the profile of the disk-like component within the cylinder is $\rm \langle n \rangle =(3.2\pm 2) \times 10^{-3}\, cm^{-3}$. 
Under the assumption that all the emission is produced by diffuse gas of thermal energy $kT=0.15$ keV, its soft X-ray (0.2-2 keV) luminosity is $\rm L_X = \epsilon_{0.2-2 keV}\cdot V_{cyl} \cdot \langle n \rangle^2 \simeq 6.4 \times 10^{39} erg\, s^{-1}$, where $\rm \epsilon_{0.2-2 keV}(kT=0.15keV,\, Z=0.1Z_\odot)=1.16 \times 10^{-15}\, erg\, cm^3\, s^{-1}$ is the soft X-ray emissivity, while the correspondent total thermal energy is then $\rm E_{th} = \langle n \rangle V kT = (1.3\pm 0.7)\times 10^{54}$ erg. \\ 
Can the luminosity $\rm L_X$ be supported by heating from supernova explosions in the solar neighbourhood?
The star formation rate (SFR) in the solar neighbourhood is found to reach a uniform density of $\rm \simeq (2.2\pm 0.8)\times 10^{-3}\, M_\odot \, yr^{-1}\, kpc^{-2}$ \citep{2021A&A...653A..63S}, thus providing $\rm SFR \simeq (6.2\pm 2.2)\times 10^{-2}\, M_\odot \, yr^{-1}$ within the closest 3 kpc. Given a star formation rate, the supernova explosion rate can be estimated trough the conversion factor $\rm \alpha = 8.8 \times 10^{-3}\, SN\, M_\odot^{-1}$ obtained for MW-like galaxies \citep{2011ApJ...738..154H, 2013ApJ...778..164A}. The rate of supernovae in the solar neighbourhood (<3 kpc) is thus $\rm \Dot{N}_{SN} = \alpha \cdot SFR = (5.4\pm 2.0) \times 10^{-4} yr^{-1}$ (i.e. 1 supernova every $\sim 1000-3000$ years). Assuming a supernova energy of $\rm E_{SN}=10^{51}$ erg spent to heat the surrounding gas, the supernovae luminosity is then $\rm L_{SN} = E_{SN} \cdot \Dot{N}_{SN} = (1.7\pm 0.6)\times 10^{40}\, erg \, s^{-1}$. The supernovae luminosity $\rm L_{SN}$ is thus comparable with the X-ray plasma luminosity $\rm L_X$, implying that the disk-like component may be powered (in part or entirely) by the current supernova rate in the stellar disk.

\subsection{Pressure balance between the disk-like component and high velocity clouds}

\begin{figure}
    \centering
    \includegraphics[width=\linewidth]{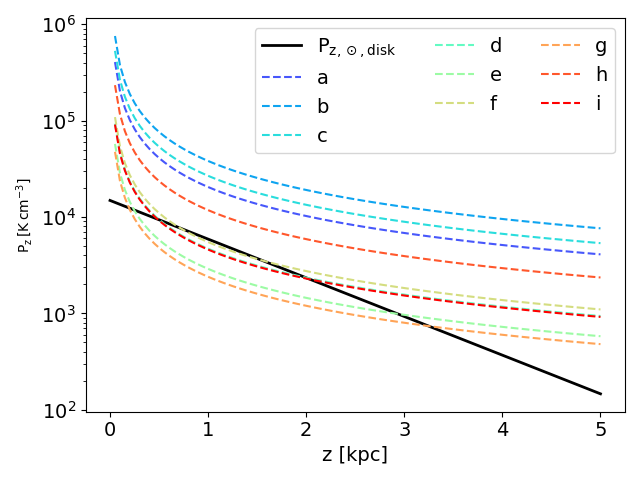}
    \caption{Vertical pressure profile $P_{z,\odot, disk}$ computed at the Sun position. Dashed lines show the pressure measurements of HVCs available in literature: from \cite{1991A&A...250..484W} MI.1 (a, [l,b]=[165,70] deg), AIV.1 and AIV.2 (b and c respectively, [l,b]=[153,40] deg), from \cite{2005ApJ...630..332F} along the line of sight of HE0226-4110 ([l,b]=[354,-66] deg) HVC.1 (d), HVC.2 (e), HVC.3 (f), HVC.4 (g) and along the line of sight of PG0953+414 ([l,b]=[180,+52] deg) HVC.1 (h) and HVC.2 (i). The pressure of the HVCs is degenerate with distance, following $\propto z^{-1}$.} 
    \label{fig:plot_HVC_P_vs_z}
\end{figure}
We construct the pressure profile of the disk-like warm-hot component as a function of height $|z|$ from the MW mid-plane at the Sun position. This pressure can be compared with the one derived for high velocity clouds (HVCs) at moderate/high Galactic latitude. HVCs show pressure gradients between the core and the outer layers requiring an external pressure support in order not to dissolve in a too short timescale \citep{1997ARA&A..35..217W}. Here we investigate if the warm-hot disk-like component can provide such a pressure support, and if equilibrium with the HVCs pressure is reached. 

We plot the HVCs pressures in Fig.~\ref{fig:plot_HVC_P_vs_z} as a function of their height from the MW midplane. Between the few HVCs with available estimates of their pressure, we select the ones at moderate/high Galactic latitudes ($|b|>40$ deg). HVCs pressure measurements in fact are degenerate with respect to the (unknown) cloud distances $d$ from us, following $P\propto d^{-1}$ \citep{1991A&A...250..484W}. For high-latitude clouds then the assumption $d\sim z$ introduces a smaller bias in the vertical pressure profile $P_z$ then for clouds at lower latitudes. We also compute the pressure profile derived for the warm-hot disk-like component of our best-fit model. The model is computed at the Sun position, following $n_\odot e^{-|z|/z_h}$. 

We find that the pressure of the disk-like gas component is in the same ballpark as the ambient pressure required for HVCs, with moderate scatter. We note that the detailed picture for each cloud may be dependent on other sources of (non-thermal) pressure support within the clouds (e.g. turbulence, magnetic, cosmic rays) that may produce the scatter in the profiles. The HVCs with highest pressure may be even found at distances larger than 5 kpc where the spherical halo pressure component starts to dominate over the disk. However, the general order of magnitude agreement in the disk suggests a physical link between the hot gas pressure and the HVCs pressure, which would not have reason to match otherwise. In addition, by assuming our model, the distance to the HVCs can be roughly derived at few kpc from the Sun, broadly agreeing with the recent constraints put on other HVCs of similar properties \citep{2022MNRAS.513.3228L}.

\subsection{Results into context}

\begin{figure}
    \centering
    \includegraphics[width=\linewidth]{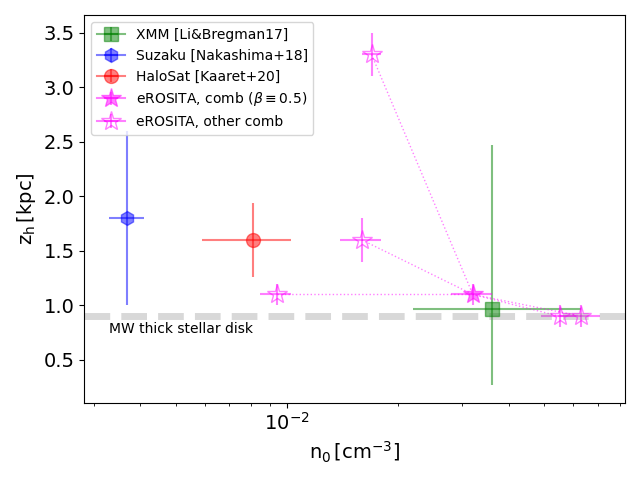}
    \caption{$z_h$ vs. $n_0$ derived by different works, as labelled. Error bars only provide statistical uncertainties quoted in the references. } \label{fig:plot_zh_vs_n0}
\end{figure}
Analyses of the O VIII line intensities for the study of the diffuse Galactic background have been already conducted using different instruments. In general, different instruments (e.g. XMM-{\it Newton}, Suzaku, HaloSat, \erosita{}) have different field of view, spectral resolution, reach different spatial resolution and cover a different total sky area. In addition, and potentially based on the information retrieved by a given instrument, different authors may adopt different assumptions on the physical properties of the hot CGM of the MW (e.g. $kT,\, Z$) when required.
Despite the differences however, a consistent picture is arising from these experiments. 

In Fig.~\ref{fig:plot_zh_vs_n0}, we show the parameters derived by similar studies, for the density profile of the disk-like component. We focus on the disk-like component as it is the one producing most of the emission attributed to the CGM X-ray emission.
We first note that thanks to the sky coverage and spatial resolution of \erosita{}, the statistical uncertainties are the smallest to date. Despite this however, physical assumptions necessarily introduce systematic biases in the results, witnessed by the significant shift of the best-fit parameters for the different models summarized in Tab.~\ref{tab:results} (empty magenta stars) with respect to the combined ($\beta\equiv 0.5$) model (filled magenta star).
Compared with the other works, we find an overall agreement on $z_h\sim 1-3$ kpc, with some preference towards the lower end of the range. The agreement likely comes from the fact that no physical assumption (i.e. $kT,\, Z$) is required to fit this parameters. Again, we note that the lower bound of a scale height of 1-3 kpc is not too far to the one estimated for the thick disk component of the MW of $z_{\rm thick}=0.9$ kpc \citep{2017MNRAS.465...76M}, as discussed above.
If the source of the disk-like emission is a truly diffuse plasma, the normalization of the profile $n_0$ really indicates a density value. Although $n_0$ shows inconsistency between experiments when only the statistical uncertainties are taken in to account, we note that assumptions on temperature and metal abundances shift $n_0$ across the parameter space. Considering that our combined ($\beta\equiv 0.5$) model assumes $kT=0.15$ keV and $Z=0.1Z_\odot$, $n_0$ is consistent with the $\sim \times 2-4$ lower values found using HaloSat \citep[$kT\simeq0.225,\, Z\equiv0.3 Z_\odot$][]{2020NatAs...4.1072K} and Suzaku \citep[$kT\simeq0.28,\, Z\equiv Z_\odot$][]{2018ApJ...862...34N}, within uncertainties, as shown by our combined+high$\epsilon$ model, holding similar assumptions. The large statistical uncertaintiy on the value estimated using XMM \citep{2017ApJ...849..105L} relates to the small sky coverage of the experiment. 
Furthermore, a minor but additional scatter across different experiments possibly relates to the different treatment of background and foreground components other than the hot CGM of the MW. 

Using a conservative approach, $n_0\simeq 1-6 \times 10^{-2}\, \rm cm^{-3}$ reasonably encompasses the actual value, although we note that, in a sense, $n_0$ has still a more geometrical meaning rather than a physical one, for at least two different reasons: 
i) our and similar experiments are currently only able to probe the CGM using average profile models, while the details at a precise location in space are expected to certainly deviate (within some scatter) from the average picture; 
ii) as $n_0$ is the extrapolation of the density profile for $(R,z) << (R_h, z_h)$, there may be no place at all holding $n=n_0$, as the physics at the Galactic Center largely deviates from the one assumed in these works.\\

\begin{figure}
    \centering
    \includegraphics[width=\linewidth]{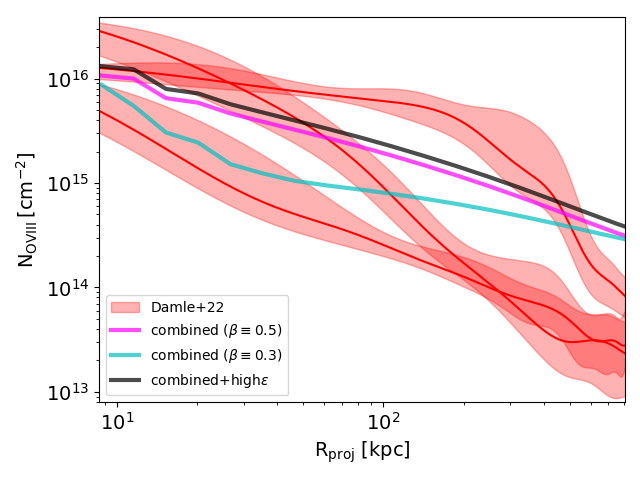}
    \caption{Column density of O VIII versus projected distance $R_{\rm proj}$, for different models (see Table~\ref{tab:results}), compared with MW profiles extracted from the HESTIA simulations of the Local Group \citep{2022MNRAS.512.3717D}.} \label{fig:plot_NO8_vs_Rproj}
\end{figure}
In Fig.~\ref{fig:plot_NO8_vs_Rproj} we compute the column density of O VIII emitting gas, as seen by an observer far away from the MW, who computes $N_{\rm OVIII}$ as a function of the projected distance from the Galactic Center. Different colors of the solid lines in Fig.~\ref{fig:plot_NO8_vs_Rproj} show the profiles for our combined ($\beta\equiv 0.5$) model (magenta) and some of the others summarized in Table~\ref{tab:results}, as labelled. In particular we show the combined ($\beta\equiv 0.3$) and combined+high$\epsilon$ to investigate how $\beta,\, kT$ and $Z$ affect $N_{\rm OVIII}$. We compare the $N_{\rm OVIII}$ profiles with prediction from the HESTIA simulations of the MW in the Local Group for different initial conditions \citep[please refer to][for details]{2022MNRAS.512.3717D}.  
Our models are all consistent with the HESTIA profiles within the first $\sim 100$ kpc. At larger distances they seem to overpredict the column density of some of the HESTIA realizations, while they become inconsistent at $R_{\rm proj} >500$ kpc, with the main difference induced by the choice of $\beta$. This distances however correspond to $\sim 2R_{\rm vir}$, where our assumptions on the physics may be broken by galaxy-galaxy interactions within the Local Group. Although this comparison does not allow us to state in favor or against some of our models, we find a general agreement between our results and the HESTIA predictions.\\

\begin{figure}
    \centering
    \includegraphics[width=\linewidth]{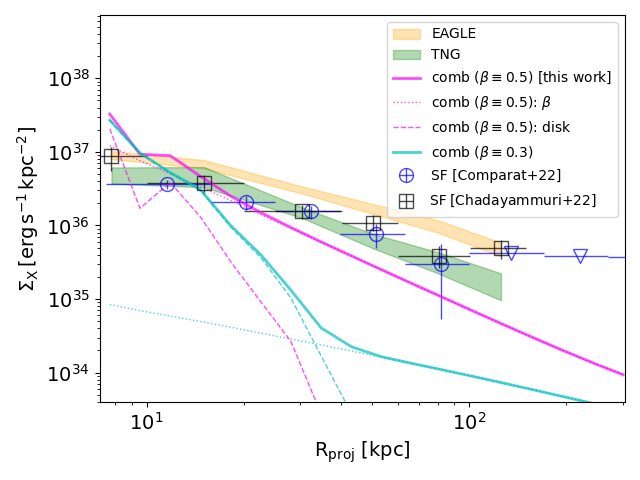}
    \caption{Projected surface brightness profile $\Sigma_X$ as a function of projected distance $R_{\rm proj}$ from the Galactic Center (as seen by an observer far away from the MW) in comparison to stacking experiments \citep{2022A&A...666A.156C, 2022ApJ...936L..15C} and simulations \citep[taken from][]{2022ApJ...936L..15C}.
    The dotted and dashed lines show the $\beta$ halo and disk-like emission components respectively. The blue triangles indicate upper limits from \citealt{2022A&A...666A.156C}.} \label{fig:plot_SigmaX_vs_Rproj}
\end{figure}
\cite{2022A&A...666A.156C} and \cite{2022ApJ...936L..15C} computed instead the average \erosita{} X-ray (0.5-2.0 keV) surface brightness profile of different populations of star-forming (SF) galaxies, by stacking cross-matched samples extracted from the eFEDS field. We show a comparison of their results with ours in Fig.~\ref{fig:plot_SigmaX_vs_Rproj}. We show the same models as in Fig.~\ref{fig:plot_NO8_vs_Rproj}, with the exception of combined+high$\epsilon$, as $kT$ and $Z$ only bias the density profile, while leaving the brightness unchanged. The combined+high$\epsilon$ model and its component would thus overlay everywhere in Fig.~\ref{fig:plot_SigmaX_vs_Rproj} to the combined ($\beta\equiv 0.5$) model (magenta).

The $\Sigma_X$ projected profiles are dominated by the disk-like ($\beta$) CGM component below (above) $R_{\rm proj} \sim 15-40$ kpc for the combined $\beta\equiv 0.3-0.5$. For the stacked data profiles, we chose the bins in stellar mass (and star formation) closer to the values inferred for the MW. Some caution has to be put in the comparison with the data for different reasons: i) in our reprojection of the MW models we did not include any absorption from the cold ISM. Any realistic galaxy seen from any angle will necessarily be highly absorbed in the central regions in the soft X-ray band. Our profiles at small $R_{\rm proj}$ are thus mostly upper limits to the actual profile ii) the data points from stacking are not PSF-subtracted and they are consistent with it up to $R_{\rm proj}\sim 80$ kpc \citep{2022A&A...666A.156C}; iii) because of the previous fact and despite the galaxy selection criteria, unresolved contribute from Active Galactic Nuclei may still be present in all the samples thus biasing high the stacking data profile for $R_{\rm proj} < 80$ kpc; iv) projection effect with close galaxies (i.e. two-halo term) is considered as a major contribute at the largest $R_{\rm proj}$. For the above reasons the profiles from stacking should be considered more like an upper limit when compared to our models.
All the models considered in Tab.~\ref{tab:results} are well within the upper limits.
In this case, both the EAGLE \citep{2015MNRAS.450.1937C, 2015MNRAS.446..521S} and TNG \citep{2018MNRAS.477..450N, 2019MNRAS.490.3196P} simulations \citep[stacked by][]{2022ApJ...936L..15C}, seems to overpredict our combined ($\beta\equiv 0.5$) model (magenta).

\section{Conclusion} \label{sec:conclusion}

In this work we analysed data from the first \erosita{} All-Sky Survey of the \erosita{}\_DE consortium. We exploited a narrow-band image ($\Delta E=80$ eV) produced in an energy range (O VIII) representative for the emission of the warm-hot component attributed to the CGM of the MW at $kT=0.15-0.23$. We retrieve the CGM emission intensity in the O VIII band by modeling and subtracting the instrumental background and CXB taking into account Galactic absorption. We fitted the CGM component using three different profiles describing the density of warm-hot gas in the MW: 
i) a spherical halo described by a $\beta$ model; 
ii) a disk-like exponential profile characterizing possible stellar feedback or population; 
iii) the linear combination of the previous two. 

We find that, in accordance to previous studies conducted with different instruments:
\begin{itemize}
    \item i) a disk-like component virtually accounts for most or all of the observed CGM emission; 
    \item ii) the inclusion of a $\beta$ halo while only slightly improving the fit of the eRASS1 data, it is also necessary to account for O VII absorption column densities observed by XMM as well as the majority of the mass; 
    \item iii) the scale height of the disk-like component in the combined model is $z_h\sim1-3$ kpc  depending on other assumptions (i.e. $\beta,\, R_h$), with a preference towards $z_h\simeq 1$ kpc; 
    \item vi) by considering diffuse warm-hot and hot plasma as the source of the disk-like component, we find that most of the emission is produced within $\sim 5$ kpc from the Sun;
    \item iv) the normalization of the spherical halo $C$ constraining the mass included within $R_{\rm vir, MW}=250$ kpc suffers of moderate systematic uncertainties mostly depending on $\beta,\, kT$ and $Z$. Thus, in contrast with previous claims, we find that the amount of baryons in the hot CGM of the MW is largely unconstrained;
    \item v) the disk-like profile and scale height is not dissimilar from updated models of MW projected mass profiles (stellar plus gas), suggesting that either some fraction of the emission attributed to the CGM may be contributed by unresolved stellar populations or that the disk-like component is a stationary (i.e. long-lived) gaseous atmosphere following the same gravitational potential as the stellar thick disk;
    \item vi) a truly diffuse nature of the disk-like component can be energetically sustained by star formation (via heating by supernovae explosions) at least locally;
    \item vii) the disk-like component of the warm-hot CGM can provide (part of) the ambient pressure support required by observations of high velocity clouds in the MW.
\end{itemize}

We also demonstrated the augmented statistical power provided by the quality and amount of the \erosita{} data. Our knowledge of the CGM properties are thus now limited by the still necessary physical assumptions on the plasma properties. These assumption will potentially be loosened by observations of the soft X-ray band with future high resolution spectrometers (e.g. XRISM, Athena), allowing to resolve individual emission lines and in turn further constraining $kT$ and $Z$ in the CGM of the MW. 

{\small \paragraph{Acknowledgements}
This work is based on data from eROSITA, the soft X-ray instrument aboard SRG, a joint Russian-German science mission supported by the Russian Space Agency (Roskosmos), in the interests of the Russian Academy of Sciences represented by its Space Research Institute (IKI), and the Deutsches Zentrum für Luft- und Raumfahrt (DLR). The SRG spacecraft was built by Lavochkin Association (NPOL) and its subcontractors, and is operated by NPOL with support from the Max Planck Institute for Extraterrestrial Physics (MPE). The development and construction of the eROSITA X-ray instrument was led by MPE, with contributions from the Dr. Karl Remeis Observatory Bamberg \& ECAP (FAU Erlangen-Nuernberg), the University of Hamburg Observatory, the Leibniz Institute for Astrophysics Potsdam (AIP), and the Institute for Astronomy and Astrophysics of the University of Tübingen, with the support of DLR and the Max Planck Society. The Argelander Institute for Astronomy of the University of Bonn and the Ludwig Maximilians Universität Munich also participated in the science preparation for eROSITA.
The eROSITA data shown here were processed using the eSASS software system developed by the German eROSITA consortium.
NL, GP and XZ acknowledge financial support from the European Research Council (ERC) under the European Union’s Horizon 2020 research and innovation program HotMilk (grant agreement No. [865637]).
GP also aknowledges support from Bando per il Finanziamento della Ricerca Fondamentale 2022 dell’Istituto Nazionale di Astrofisica (INAF): GO Large program. 
The authors thank Mattia Sormani and Shifra Mandel for fruitful discussion and help. }

\bibliographystyle{aa}
\bibliography{aa} 



\appendix

\section{X-ray absorption} \label{sec:appendix_A}

\begin{figure}
    \centering
    \includegraphics[width=\linewidth]{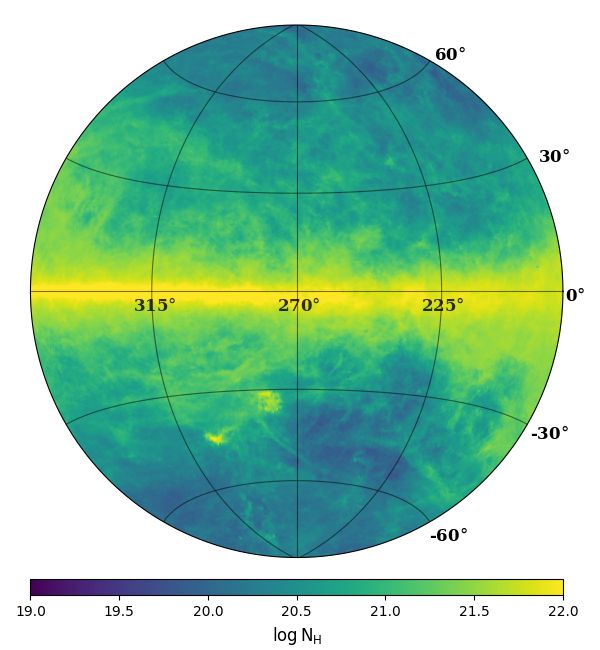}
    \caption{Total hydrogen column density $\rm N_H$ map of the foreground absorption.
    }
    \label{fig:hmap_NHtot_n128_id9100}
\end{figure}
We assume the neutral atomic hydrogen column densities $N_{\rm HI}$ from the full sky {\small HI4PI} radio survey \citep{2016A&A...594A.116H}. In general, the content of atomic hydrogen $N_{\rm HI}$ differs from the total content of hydrogen in both atomic and molecular form $N_H$. 
We compute the total hydrogen column density $N_H$ as 
\begin{align}
    N_H & = N_{\rm HI} + 2 N_{\rm H_2} \\
    N_{\rm H_2} & = 7.2 \times 10^{20} \left[ 1 - \exp \left( \frac{- N_{\rm HI} E_{\rm B-V} }{ 3.0 \times 10^{20} } \right) \right]^{1.1}    
\end{align}
where $N_{\rm H_2}$ is the molecular hydrogen column density in cm$^{-2}$ and $E_{\rm B-V}$ is the optical reddening caused by dust extinction \citep[][eq. 7]{Willingale.ea:13}. The optical reddening have been taken from the \cite{2016A&A...596A.109P} results, extracted using to the {\tt dustmaps} tool of \cite{2018JOSS....3..695M}.
The computed total hydrogen column density map $\rm N_H$ is shown in Fig.~\ref{fig:hmap_NHtot_n128_id9100}. At high absolute Galactic latitudes, the Northern hemisphere shows on average a larger column density with respect to the Southern hemisphere.
The total hydrogen column density have also been computed through a direct correlation with the optical extinction in the V-band $A_v$ as $N_H = 2.04 \times 10^{21} A_v$ \citep{2017MNRAS.471.3494Z}, where $A_v = R_v E_{\rm B-V}$ with $R_v = 3.1$ the  total-to-selective extinction ratio.
The two methods provide consistent $N_H$ values, with the largest discrepancies produced only for large $N_{\rm HI}$ lines of sight, mainly involving very low Galactic latitudes.
In fact, we note that at high enough Galactic latitudes, where we focus our analysis, the relatively low column densities ($N_{\rm HI}<10^{21}\, \rm cm^{-2}$) limit the presence of molecular gas not probed by HI surveys and makes the two different $N_H$ estimates consistent with each other within a few percent.
We assume the CGM and CXB emission components to be located behind all of the absorbing layers, thus accounting for the whole $N_H$ value, for all lines of sight.

\section{Local Hot Bubble}

The LHB emission becomes increasingly important towards softer energies. 
This fact is supported from both spectral analysis (Fig.~5 in \citealt{2022arXiv221003133P}) and from the morphology of the soft emission, highly resembling that of the 1/4 keV ROSAT map, thought to be dominated by such component \citep{1997ApJ...485..125S}. 
The latter argument in particular, considering the different background particle environments of \erosita \, and ROSAT \citep[][]{1997ApJ...485..125S, 2021A&A...647A...1P} implies the importance of LHB photons with respect to the instrumental particle background in the soft band.
A detailed modeling of the LHB emission measure has been provided by \cite{2017ApJ...834...33L} who quantified and removed from the ROSAT data the contribution of the solar wind charge exchange (SWCX) spurious component, thanks to data collected with a dedicated rocket mission. By assuming thermal ionization equilibrium, the authors provided an emission measure map of the LHB, found to be at a remarkably homogeneous temperature of $kT = 0.097 \pm 0.019$ keV across the sky.
Although their map derived from the ROSAT data has lower angular resolution than \erosita,\, it is the most robust modeling of the LHB currently available thanks to the large effective area of ROSAT at very soft energies $E\sim 0.1$ keV, where the emission from all the diffuse components become heavily absorbed, except from the LHB. 
By assuming solar metallicities \citep{2022arXiv221003133P}, we then reconstructed the expected LHB emission in all bands, estimating median contributes of $1.20 \times 10^{-2}\,\rm ph\, s^{-1} \deg^{-2}$ in the O VIII energy band (i.e. $\sim 1.3\%$ of the line intensity). The LHB thus only provides a very small fraction of the counts in the O VIII energy band. 
We attribute a fractional error of 8\% to the LHB count rates. This includes possible systematic variations of the LHB temperature estimation \cite{2017ApJ...834...33L}.

\section{Solar wind charge exchange} \label{appendix:swcx}

The Solar wind is known to be composed of ionised particles which can interact with the neutral ISM. The interaction between the ionized and neutral material through the exchange of electrons was first discovered to produce soft X-rays in the tails of comets \citep{1997GeoRL..24..105C} and at the interface between the heliosphere and the ISM \citep{1998LNP...506..121C}.
The ISM in fact constantly passes through the heliosphere due to its relative motion with respect to the Sun. Emission lines consistent with charge exchange from the solar wind have been detected by XMM-{\it Newton} \citep{2004ApJ...610.1182S} and their contribution to the 1/4 KeV ROSAT map has been found to be important and widespread across the sky \citep{2004A&A...418..143L}.
Due to the variability and anisotropy of the Solar wind, the emission of the Solar Wind Charge eXchange (SWCX) is also found to be variable in both time and direction \citep{2019A&ARv..27....1K}.
An additional component of the SWCX can be produced at the interaction region between the Solar wind and the Earth's exosphere. However, thanks to the position of \erosita\, in L2, this component is expected to be negligible with respect to the heliospheric one within our data, while it is potentially relevant for instruments operating at low Earth orbits, such as XMM{\it -Newton} or {\it Chandra} \citep{2004ApJ...610.1182S}.

The intensity of the eRASS1 data is found to be consistent with a negligible variation with respect to the eRASS2 data (\citealt{2022arXiv221003133P}, Dennerl et al. in preparation). This implies that if the eRASS1 data are affected by SWCX emission, this component is rather stable over a $\sim$1 yr time scale, at odds with the SWCX properties.
In any case, the SWCX intensity has been constrained in the eRASS1 data in the direction of dark clouds \citep{Yeung2023} to amount to $< 10\%$ of the median observed O VIII intensity. Given the direction towards the GC and close to the ecliptic plane, both directions towards which the SWCX is thought to be highest, we can consider this value as an upper limit to the SWCX contribution in the O VIII band.
In all the other lines the SWCX emission is negligible.

Accounting for the SWCX emission in spectral analysis shifts the temperature of the warm-hot component from T=0.15 keV to 0.17 keV \citep{2022arXiv221003133P}. 
In the combined ($\beta\equiv 0.5$) model for the eRASS1 data, we consider the SWCX emission as negligible. However, we investigate what systematic this component may introduce in our results by including it and considering T=0.17 keV as one of the hot CGM phases temperature, in Sec.~\ref{sec:discussion}.

\section{Additional figures} \label{appendix:other}

\begin{figure}
    \centering
    \includegraphics[width=\linewidth]{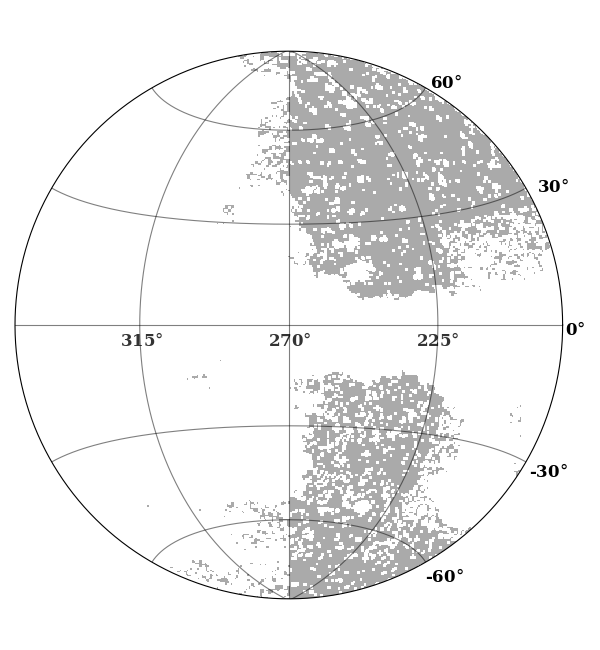}
    \caption{Data selection mask, resulting from a combination of the pixels signal to noise ratio $<2\sigma_b$ and total hydrogen column density $\rm N_H<1.6\times 10^{21}\, cm^{-2}$.}
    \label{fig:hmap_valid_n128_id7}
\end{figure}
\begin{figure*}
    \includegraphics[width=\textwidth]{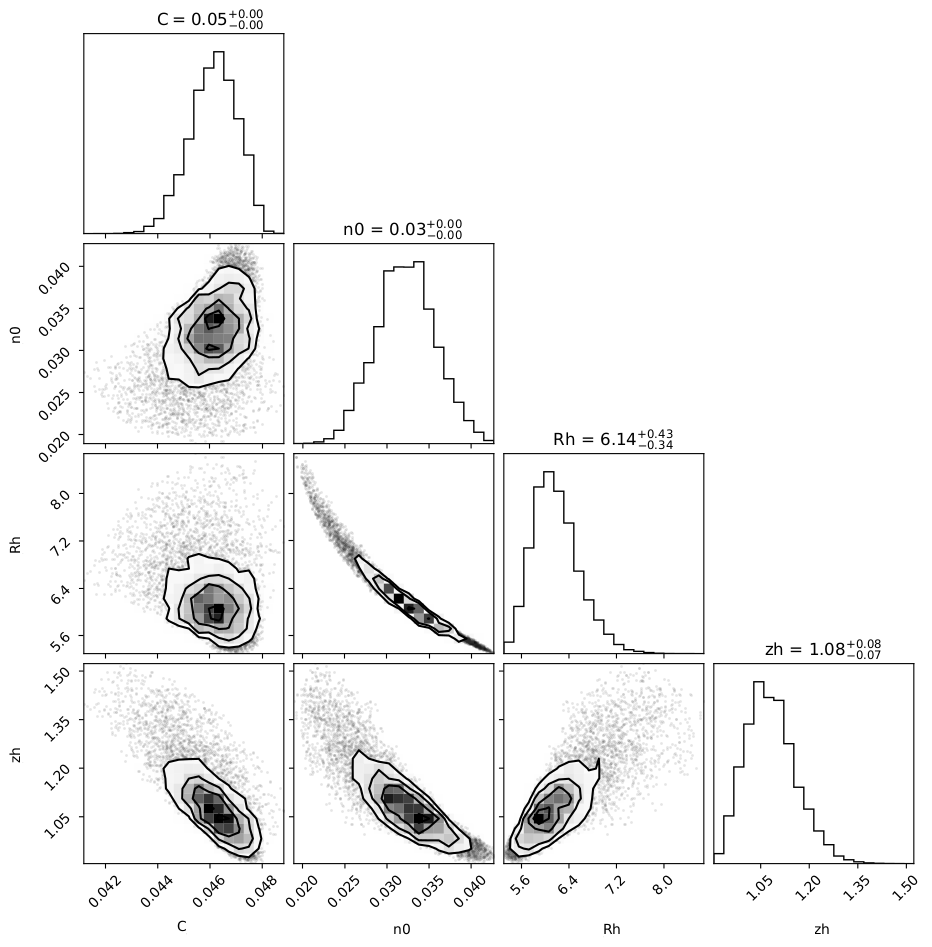}
    \caption{Corner plot for the parameters of the combined ($\beta\equiv 0.5$) model fit using {\tt ultranest}.}
    \label{fig:X18O8B05_n128_id7_corner}
\end{figure*}
%


\label{lastpage}
\end{document}